\documentclass[12pt,journal,onecolumn]{IEEEtran}
\usepackage{epsfig,rotating,setspace,latexsym,amsmath,epsf,amssymb,amsfonts,bm,theorem,subfigure,epstopdf}
\usepackage{cite,authblk}
\usepackage{algorithmic,algorithm}
\usepackage{setspace}
\usepackage{color}
\usepackage{bbm}
\newtheorem{lemma}{Lemma}

\newenvironment{Proof}[1]{\medskip\par\noindent{\bf Proof:\,}\,#1}{{\mbox{\,$\blacksquare$}\par}}

\allowdisplaybreaks

\begin{document}

\title{Using Timeliness in Tracking Infections  \thanks{This work was supported by NSF Grants CCF 17-13977 and ECCS 18-07348.}}

\author{Melih Bastopcu$^{1}$ \qquad Sennur Ulukus$^{2}$\\
	\normalsize $^{1}$Coordinated Science Laboratory, University of Illinois Urbana-Champaign, IL\\
	\normalsize $^{2}$Department of Electrical and Computer Engineering, University of Maryland, MD\\
	\normalsize  \emph{bastopcu@illinois.edu} \qquad \emph{ulukus@umd.edu}}

\maketitle

\vspace{-1.5cm}

\begin{abstract}
We consider real-time timely tracking of infection status (e.g., covid-19) of individuals in a population. In this work, a health care provider wants to detect infected people as well as people who have recovered from the disease as quickly as possible. In order to measure the timeliness of the tracking process, we use the long-term average difference between the actual infection status of the people and their real-time estimate by the health care provider based on the most recent test results. We first find an analytical expression for this average difference for given test rates, infection rates and recovery rates of people. Next, we propose an alternating minimization based algorithm to find the test rates that minimize the average difference. We observe that if the total test rate is limited, instead of testing all members of the population equally, only a portion of the population may be tested in unequal rates calculated based on their infection and recovery rates. Next, we characterize the average difference when the test measurements are erroneous (i.e., noisy). Further, we consider the case where the infection status of individuals may be dependent, which happens when an infected person spreads the disease to another person if they are not detected and isolated by the health care provider. In addition, we consider an age of incorrect information based error metric where the staleness metric increases linearly over time as long as the health care provider does not detect the changes in the infection status of the people. Through extensive numerical results, we observe that increasing the total test rate helps track the infection status better. In addition, an increased population size increases diversity of people with different infection and recovery rates, which may be exploited to spend testing capacity more efficiently, thereby improving the system performance. Depending on the health care provider's preferences, test rate allocation can be adjusted to detect either the infected people or the recovered people more quickly. When there are errors in the tests, in order to combat the test errors, it may be more advantageous for the health care provider not to test everyone, and instead, apply more tests to a selected portion of the population. In the case of people with dependent infection status, as we increase the total test rate, the health care provider detects the infected people more quickly, and thus, the average time that a person stays infected decreases. Finally, the error metric needs to be chosen carefully to meet the priorities of the health care provider, as the error metric used greatly influences who will be tested and at what test rates.
\end{abstract}

\section{Introduction}
We consider the problem of timely tracking of an infectious disease, e.g., covid-19, in a population of $n$ people. In this problem, a health care provider wants to detect infected people as quickly as possible in order to take precautions such as isolating them from the rest of the population. The health care provider also wants to detect people who have recovered from the disease as soon as possible since these people need to return to work which is especially critical in sectors such as education, food retail, public transportation, etc. Ideally, the health care provider should test all people all the time. However, as the total test rate is limited, the question is how frequently the health care provider should apply tests on these people when their infection and recovery rates are known. In a broader sense, this problem is related to timely tracking of multiple processes in a resource-constrained setting where each process takes binary values of $0$ and $1$ with different change rates.

\begin{figure}[t]
	\centerline{\includegraphics[width=0.8\columnwidth]{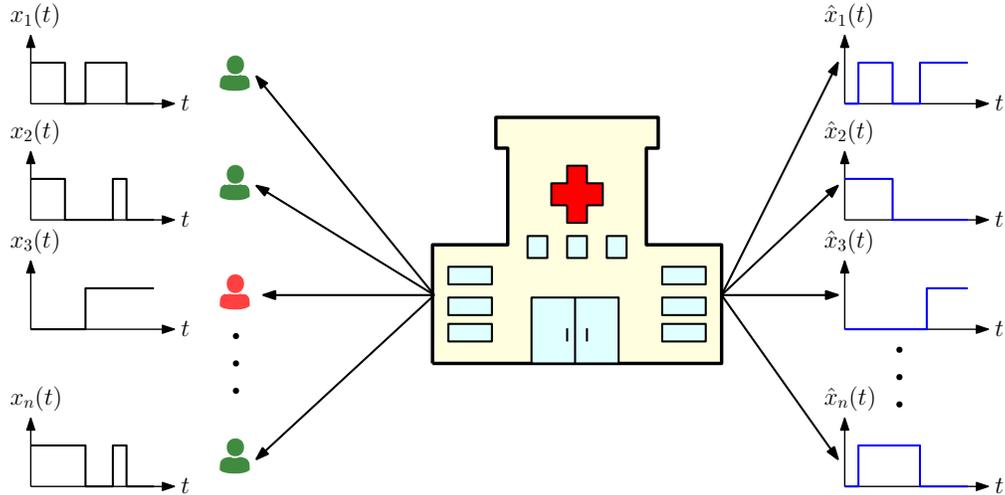}}
	\caption{System model. There are $n$ people whose infection status are given by $x_i(t)$. The health care provider applies tests on these people. Based on the test results, estimations for the infection status $\hat{x}_i(t)$ are generated. Infected people are shown in red and healthy people are shown in green.}
	\label{Fig:system_model}
\end{figure}

Recent studies have shown that people who recovered from infectious diseases such as covid-19 can be reinfected. Furthermore, the recovery times of individuals from the disease may vary significantly. For these reasons, in this problem, the $i$th person gets infected with rate $\lambda_i$ which is independent of the others. Similarly, the $i$th person recovers from the disease with rate $\mu_i$. We note that the index $i$ may represent a specific individual or a group of individuals that have common features such as age, gender, profession. Depending on the demographics, coefficients $\lambda_i$ and $\mu_i$ may be statistically known by the health care provider. We denote the infection status of the $i$th person as $x_i(t)$ (shown with the black curves on the left in Fig.~\ref{Fig:system_model}) which takes the value 1 when the person is infected and the value 0 when the person is healthy. The health care provider applies tests to people marked as healthy with rate $s_i$ and to people marked as infected with rate $c_i$. Based on the test results, the health care provider forms an estimate for the infection status of the $i$th person denoted by $\hat{x}_{i}(t)$ (shown with the blue curves on the right in Fig.~\ref{Fig:system_model}) which takes the value 1 when the most recent test result is positive and the value 0 when it is negative. 

We measure the timeliness of the tracking process by the difference between the actual infection status of people and the real-time estimate of the health care provider which is based on the most recent test results. The difference can occur in two different cases: i) when the person is sick ($x_i(t) =1$) and the health care provider maps this person as healthy ($\hat{x}_i(t) =0$), and ii) when the person recovers from the disease ($x_i(t) =0$) but the health care provider still considers this person as infected ($\hat{x}_i(t) =1$). The former case represents the error due to late detection of infected people, while the latter case represents the error due to late detection of healed people. Depending on the health care provider's preferences, detecting infected people may be more important than detecting recovered people (controlling infection), or the other way around (returning people to workforce).  

Age of information has been proposed to measure timeliness of information in communication systems, and has been studied in the context of queueing systems \cite{Kaul12a, Kadota18a, Kam16b, Sun17a, Najm18b, Soysal19, Buyukates20e, Yates20}, multi-hop and multi-cast networks \cite{Talak17, Tripathi17, Bedewy17b, Zhong18b, Buyukates18, Buyukates18b, Buyukates19, Krishnan19, Farazi19}, social networks \cite{Ioannidis09}, timely remote estimation of random processes \cite{Wang19a, Sun17b, Sun18b, Chakravorty18, Kam20a, arafa2020, Bastopcu20a}, energy harvesting systems \cite{ Bacinoglu18, Baknina18b, Baknina18a, Wu18, Feng18a, Feng18c, Arafa18a, Arafa18b, Arafa18c, Arafa18f, Arafa19e, Arafa22, Farazi18, Leng19, Chen19}, wireless fading channels \cite{Bhat19, Ostman19}, scheduling in networks \cite{ Bastopcu18, Bastopcu19b, Buyukates18c, Buyukates19b, Buyukates21d, Zhong18a, Rajaraman18, Liu19, Maatouk20a, Uysal21, Ayan19, bastopcu20f, Banerjee20}, lossless and lossy source and channel coding \cite{Zhong16, Zhong18f, Mayekar18, Mayekar20, MelihBatu1, MelihBatu2, MelihBatu4, Ramirez19, Arafa19b, arafa21a, Bastopcu20}, vehicular, IoT and UAV systems \cite{ Elmagid18, Liu18, Elmagid19c, Alabbasi20}, caching systems \cite{Gao12, Yates17b, Kam17b,  Zhang18, Tang19, Zhong18c, Yang19a, Bastopcu20c, Bastopcu21a, bastopcu20e, kaswan2021, Gu21}, computation-intensive systems \cite{Kuang19, Gong19, Zou19b, Bastopcu19, Bastopcu20b, Buyukates20b, Buyukates19c, Zou19a}, learning systems \cite{Buyukates21c, Ozfatura20a, Ceran21}, gossip networks \cite{yates2021age, Buyukates21a, Bastopcu21c, Kaswan22} and so on. A more detailed review of the age of information literature can be found in references \cite{Kosta17c, SunSurvey, Yates20a}. Most relevant to our work, the real-time timely estimation of a single and multiple counting processes \cite{Wang19a, Bastopcu20a}, a Wiener process \cite{Sun17b}, a random walk process \cite{Yun18}, binary and multiple states Markov sources \cite{Kam20a, Maatouk20a, Maatouk20b} have been studied. The work that is closest to our work is reference \cite{Kam20a} where the remote estimation of a symmetric binary Markov source is studied in a time-slotted system by finding the optimal sampling policies via formulating a Markov Decision Process (MDP) for real-time error, AoI and AoII metrics. Different from \cite{Kam20a}, in our work, we consider real-time timely estimation of multiple non-symmetric binary sources for a continuous time system. In our work, the sampler (health care provider) does not know the states of the sources (infection status of people), and thus, takes the samples (applies medical tests) randomly (exponential random variables) with fixed rates. Thus, we optimize the test rates of people to minimize the real-time estimation error.          

In this paper, we consider the real-time timely tracking of infection status of $n$ people. We first find an analytical expression for the long-term average difference between the actual infection status of people and the estimate of the health care provider based on test results. Then, we propose an alternating minimization based algorithm to find the test rates $s_i$ and $c_i$ for all people. We observe that if the total test rate is limited, we may not apply tests on all people equally. Next, we provide an alternative method to characterize the average difference, by finding the steady state of a Markov chain defined by $(x_i(t),\hat{x}_i(t))$. By using this alternative method, we determine the average estimation error when there are errors in the test measurements expressed by a false positive rate $p$ and a false negative rate $q$. Next, we consider the infection status of two people where an infected person may spread the disease to another person if the infection has not been detected by the health care provider to isolate the infected person. Finally, we consider an age of incorrect information based error metric where the estimation error increases linearly over time when the health care provider has not detected the changes in the infection status of the people. 

Through extensive numerical results, we observe that increasing the total test rate helps track the infection status of people better, and increasing the size of the population increases diversity which may be exploited to improve the performance. Depending on the health care provider's priorities, we can allocate more tests to people marked as healthy to detect the infections more quickly or to people marked as infected to detect the recoveries more quickly. In order to combat the test errors, the health care provider may prefer to apply tests to only a selected portion of the population with higher test rates. When the infection status of a person depends on that of another person, the average time that a person remains infected can be reduced by increasing the total test rate as it helps to detect the infected people more quickly. Finally, we observe that depending on the error metric used, the test rate distribution among the population differs greatly, and thus, we should choose an error metric that aligns with the priorities of the health care provider.        

\section{System Model} \label{sect:system_model}
\begin{figure}[t]
\begin{center}
\subfigure[]
{\includegraphics[width=0.49\linewidth]{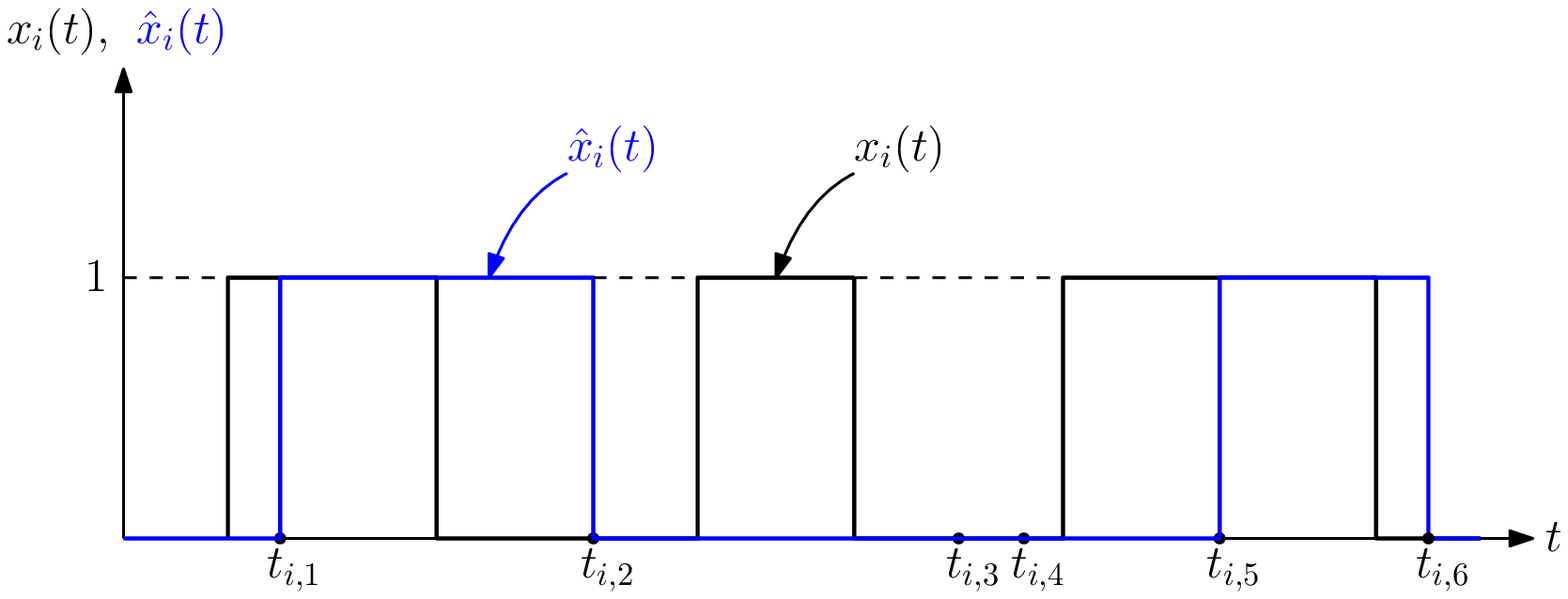}} 
\subfigure[]
{\includegraphics[width=0.49\linewidth]{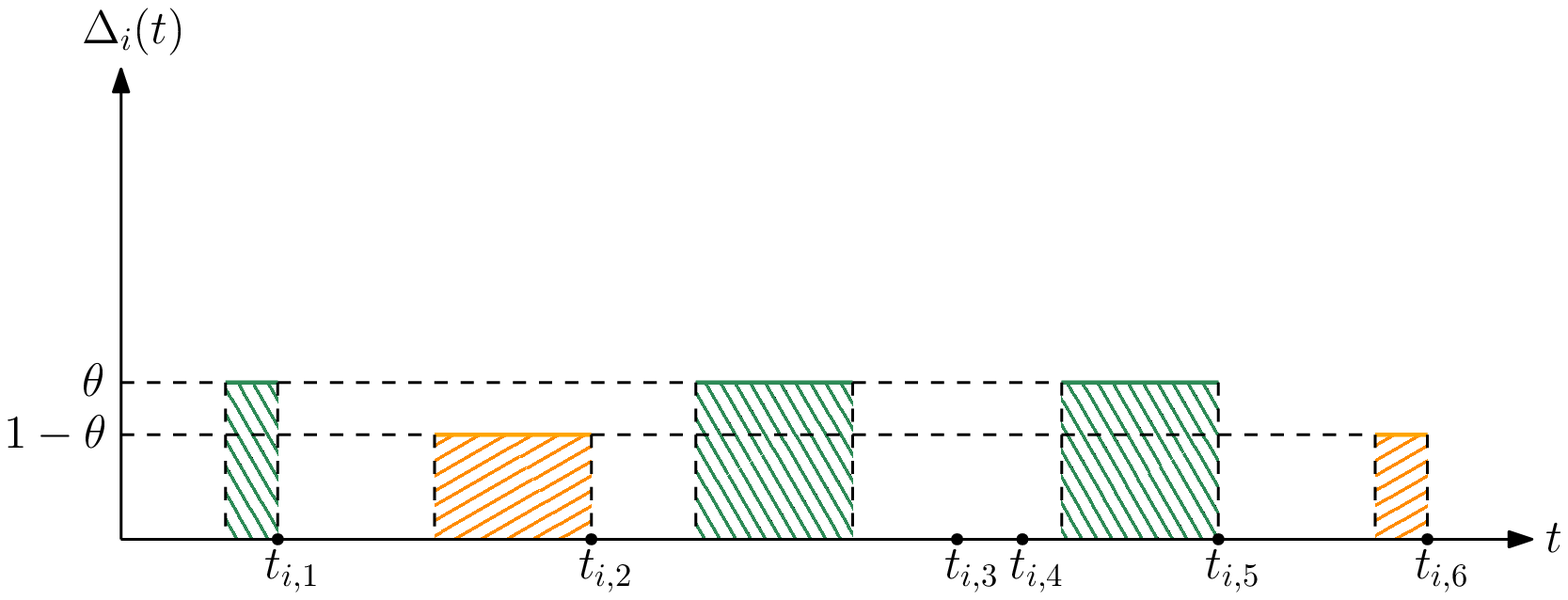}}
\end{center}
\caption{(a) A sample evolution of $x_i(t)$ and $\hat{x}_i(t)$, and (b) the corresponding $\Delta_i(t)$ in (\ref{delta_t}). Green areas correspond to the error caused by $\Delta_{i1}(t)$ in (\ref{delta_i1}). Orange areas correspond to the error caused by $\Delta_{i2}(t)$ in (\ref{delta_i2}).}
\label{fig:model}
\end{figure}

We consider a population of $n$ people. We denote the infection status of the $i$th person at time $t$ as $x_i(t)$ (black curve in Fig.~\ref{fig:model}(a)) which takes binary values $0$ or $1$ as follows,
\begin{align}
x_{i}(t) = \begin{cases} 
1, & \text{if the $i$th person is infected at time $t$}, \\
0, & \text{otherwise}.
\end{cases}
\end{align}

In this paper, we consider a model where each person can be infected multiple times after recovering from the disease. We denote the time interval that the $i$th person stays healthy for the $j$th time as $W_i(j)$ which is exponentially distributed with rate $\lambda_i$. We denote the recovery time for the $i$th person after infected with the virus for the $j$th time as $R_i(j)$ which is exponentially distributed with rate $\mu_i$.

A health care provider wants to track the infection status of each person. Based on the test results at times $t_{i,\ell}$, the health care provider generates an estimate for the status of the $i$th person denoted as $\hat{x}_i(t)$ (blue curve in Fig.~\ref{fig:model}(a)) by 
\begin{align}
\hat{x}_{i}(t) = x_i(t_{i,\ell}), \quad t_{i,\ell}\leq t<t_{i,\ell+1}.
\end{align}
When $\hat{x}_{i}(t)$ is 1, the health care provider applies the next test to the $i$th person after an exponentially distributed time with rate $c_i$. When $\hat{x}_{i}(t)$ is 0, the next test is applied to the $i$th person after an exponentially distributed time with rate $s_i$. 

An estimation error happens when the actual infection status of the $i$th person, $x_i(t)$, is different than the estimate of the health care provider, $\hat{x}_{i}(t)$, at time $t$. This could happen in two ways: when $x_i(t) = 1$ and $\hat{x}_{i}(t) = 0$, i.e., when the $i$th person is sick, but it has not been detected by the health care provider, and when $x_i(t) = 0$ and $\hat{x}_{i}(t) = 1$, i.e., when the $i$th person has recovered, but the health care provider does not know that the $i$th person has recovered. 

We denote the error caused by the former case, i.e., when $x_i(t) = 1$ and $\hat{x}_{i}(t) = 0$, by $\Delta_{i1}(t)$ (green areas in Fig.~\ref{fig:model}(b)), 
\begin{align}\label{delta_i1}
\Delta_{i1}(t) = \max \{x_i(t)- \hat{x}_{i}(t), 0\},
\end{align}
and we denote the error caused by the latter case, i.e., when $x_i(t) = 0$ and $\hat{x}_{i}(t) = 1$, by $\Delta_{i2}(t)$ (orange areas in Fig.~\ref{fig:model}(b)), 
\begin{align}\label{delta_i2}
\Delta_{i2}(t) = \max \{\hat{x}_{i}(t)-x_i(t), 0\}.
\end{align}
Then, the total estimation error for the $i$th person $\Delta_i(t)$ is 
\begin{align}\label{delta_t}
\Delta_{i}(t) = \theta \Delta_{i1}(t) + (1-\theta)\Delta_{i2}(t), 
\end{align}
where $\theta$ is the importance factor in $[0,1]$. A large $\theta$ gives more importance to the detection of infected people, and a small $\theta$ gives more importance to the detection of recovered people. 

We define the long-term weighted average difference between $x_i(t)$ and $\hat{x}_i(t)$ as  
\begin{align}\label{long_term}
   \Delta_{i} = \lim_{T\to\infty} \frac{1}{T}\int_0^T \Delta_{i}(t)dt.
\end{align}
Then, the overall average difference of all people $\Delta$ is
\begin{align}\label{total_age}
   \Delta = \frac{1}{n} \sum_{i=1}^{n} \Delta_{i}.
\end{align}

Our aim is to track the infection status of all people. Due to limited resources, there is a total test rate constraint $\sum_{i=1}^{n}s_i+\sum_{i=1}^{n}c_i\leq C$. Thus, our aim is to find the optimal test rates $s_i$ and $c_i$ to minimize $\Delta$ in (\ref{total_age}) while satisfying this total test rate constraint. We formulate the following problem,
\begin{align}
\label{problem1}
\min_{\{s_i, c_i \}}  \quad &  \Delta \nonumber \\
\mbox{s.t.} \quad & \sum_{i=1}^{n} s_i + \sum_{i=1}^{n} c_i\leq C \nonumber \\
\quad & s_i\geq 0, \quad c_i\geq 0,\quad i=1,\dots,n.
\end{align} 
In the next section, we find the total average difference $\Delta$. 

\begin{figure}[t]
\begin{center}
\subfigure[]
{\includegraphics[width=0.49\linewidth]{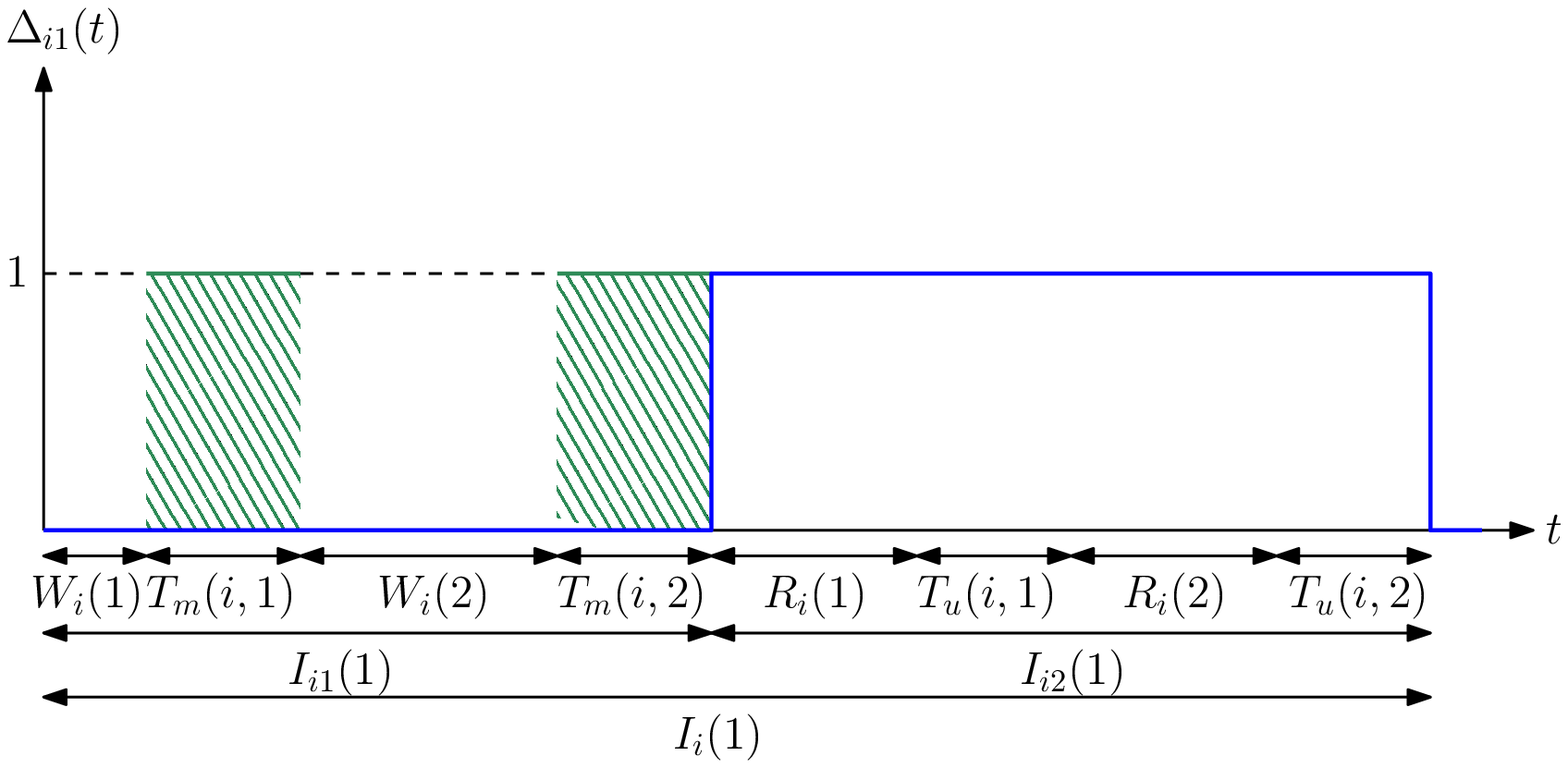}} 
\subfigure[]
{\includegraphics[width=0.49\linewidth]{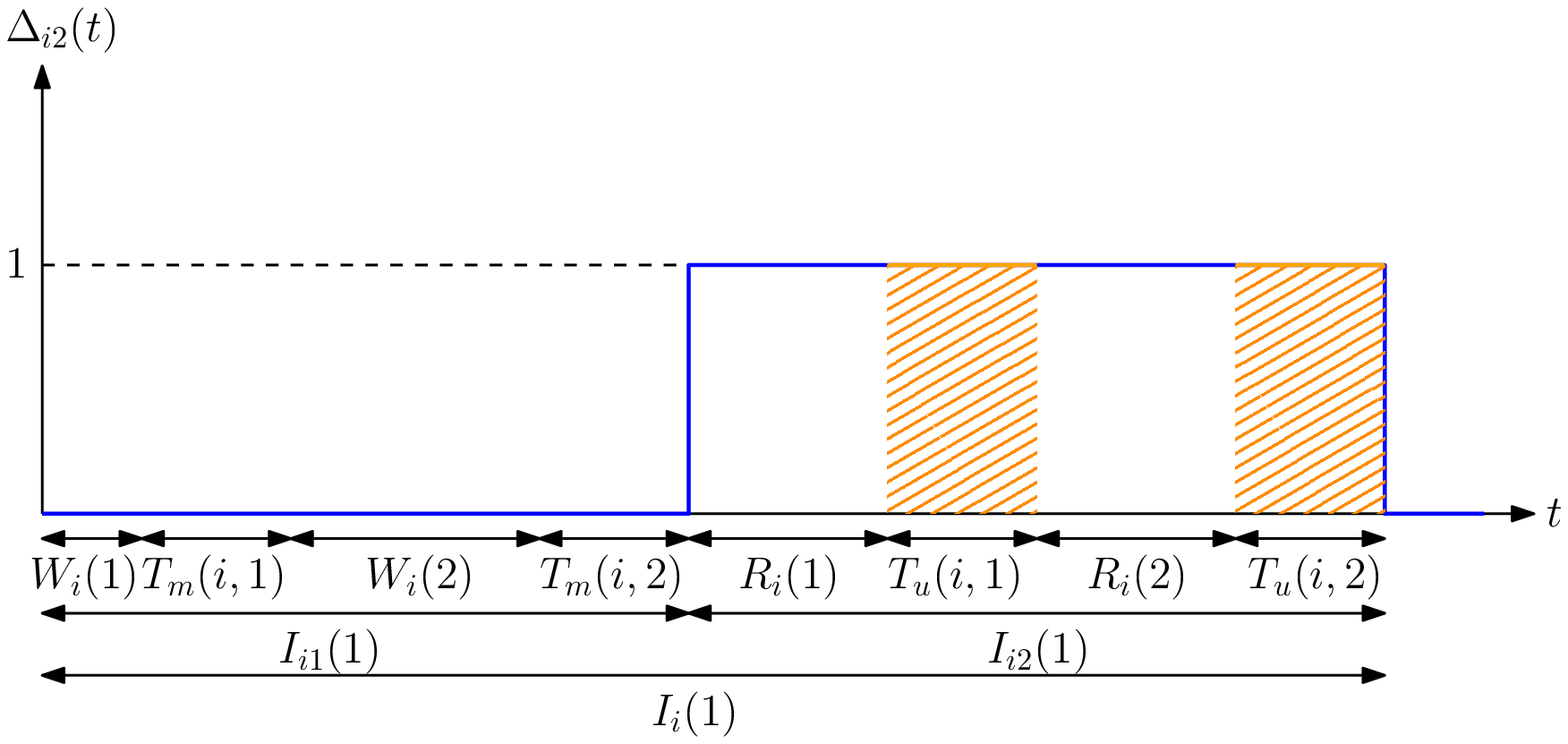}}
\end{center}
\caption{A sample evolution of (a) $\Delta_{i1}(t)$, and (b) $\Delta_{i2}(t)$ in a typical cycle.}
\label{fig:error}
\end{figure}

\section{Average Difference Analysis} \label{Sec:Average_difference} 
We first find analytical expressions for $\Delta_{i1}(t)$ in (\ref{delta_i1}) and $\Delta_{i2}(t)$ in (\ref{delta_i2}) when $s_i>0$ and $c_i>0$. We note that $\Delta_{i1}(t)$ can be equal to $1$ when $\hat{x}_i(t) = 0$ and is always equal to $0$ when $\hat{x}_i(t) = 1$. Assume that at time $0$, both $x_i(0)$ and $\hat{x}_i(0)$ are 0. After an exponentially distributed time with rate $\lambda_i$, which is denoted by $W_i$, the $i$th person is infected, and thus $x_i(t)$ becomes $1$. At that time, since $\hat{x}_i(t) = 0$, $\Delta_{i1}(t)$ becomes $1$. $\Delta_{i1}(t)$ will be equal to 0 again either when the $i$th person recovers from the disease which happens after $R_i$ which is exponentially distributed with rate $\mu_i$ or when the health care provider performs a test on the $i$th person after $D_i$ which is exponentially distributed with rate $s_i$. We define $T_m(i)$ as the earliest time at which one of these two cases happens, i.e., $T_m(i) = \min \{R_i,D_i \}$. We note that $T_m(i)$ is also exponentially distributed with rate $ \mu_i+s_i$, and we have $\mathbb{P} (T_m(i) = R_i) = \frac{\mu_i}{\mu_i+s_i}$ and $\mathbb{P} (T_m(i) = D_i) = \frac{s_i}{\mu_i+s_i}$. If the $i$th person recovers from the disease before testing, we return to the initial case where both $x_i(t)$ and $\hat{x}_i(t)$ are equal to $0$ again. In this case, this cycle repeats itself, i.e., the $i$th person becomes sick again after $W_i$ and $\Delta_{i1}(t)$ remains as 1 until either the person recovers or the health care provider performs a test which takes another $T_m(i)$ duration. If the health care provider performs a test before the person recovers, then $\hat{x}_i(t)$ becomes $1$. We denote the time interval for which $\hat{x}_i(t)$ stays at 0 as $I_{i1}$ which is given by 
\begin{align}
    I_{i1} = \sum_{\ell = 1}^{K_1} T_m(i,\ell)+ W_i(\ell), 
\end{align}
where $K_1$ is geometric with rate $\mathbb{P} (T_m(i) = D_i) = \frac{s_i}{\mu_i+s_i}$. Due to \cite[Prob. 9.4.1]{Yates14}, $\sum_{\ell = 1}^{K_1} T_m(i,\ell)$ and $\sum_{\ell = 1}^{K_1} W_i(\ell)$ are exponentially distributed with rates $s_i$ and $\frac{\lambda_i s_i}{\mu_i+s_i}$, respectively. As $\mathbb{E}[I_{i1}] = \mathbb{E}[\sum_{\ell = 1}^{K_1} T_m(i,\ell)]+\mathbb{E}[\sum_{\ell = 1}^{K_1} W_i(\ell)]$, we have
\begin{align}\label{I_1}
\mathbb{E}[I_{i1}] = \frac{1}{s_i}+\frac{s_i+\mu_i}{s_i\lambda_i }.
\end{align}

When $\hat{x}_i(t) =1$, the health care provider marks the $i$th person as infected. The $i$th person recovers from the virus after $R_i$. After the $i$th person recovers, either the health care provider performs a test after $Z_i$  which is exponentially distributed with rate $c_i$ or the $i$th person is reinfected with the virus which takes $W_i$ time. We define $T_u(i)$ as the earliest time at which one of these two cases happens, i.e., $T_u(i) = \min \{W_i, Z_i\}$. Similarly, we note that $T_u(i)$ is exponentially distributed with rate $\lambda_i+c_i$, and we have $\mathbb{P} (T_u(i) = W_i) = \frac{\lambda_i}{\lambda_i+c_i}$ and $\mathbb{P} (T_u(i) = Z_i) = \frac{c_i}{\lambda_i+c_i}$. If the person is reinfected with the virus before a test is applied, this cycle repeats itself, i.e., the $i$th person recovers after another $R_i$, and then either a test is applied to the $i$th person, or the person is infected again which takes another $T_u(i)$. If the health care provider performs a test to the $i$th person before the person is reinfected, the health care provider marks the $i$th person as healthy again, i.e., $\hat{x}_i(t)$ becomes 0. We denote the time interval that $\hat{x}_i(t)$ is equal to 1 as $I_{i2}$ which is given by
\begin{align}
    I_{i2} = \sum_{\ell = 1}^{K_2} T_u(i,\ell)+ R_i(\ell), 
\end{align}
where $K_2$ is geometric with rate $\mathbb{P} (T_u(i) = Z_i) = \frac{c_i}{\lambda_i+c_i}$. Similarly, $\sum_{\ell = 1}^{K_2} T_u(i,\ell)$ and $\sum_{\ell = 1}^{K_2} R_i(\ell)$ are exponentially distributed with rates $c_i$ and $\frac{c_i\mu_i}{\lambda_i+c_i}$, respectively. As $\mathbb{E}[I_{i2}] = \mathbb{E}[\sum_{\ell = 1}^{K_2} T_u(i,\ell)]+\mathbb{E}[\sum_{\ell = 1}^{K_2} R_i(\ell)]$, we have
\begin{align}\label{I_2}
    \mathbb{E}[I_{i2}] = \frac{1}{c_i}+\frac{c_i+\lambda_i}{c_i\mu_i }.
\end{align}

We denote the time interval between the $j$th and $(j+1)$th times that $\hat{x}_i(t)$ changes from 1 to 0 as the $j$th cycle $I_i(j)$ where $I_i(j) =I_{i1}(j)+ I_{i2}(j)$. We note that $\Delta_{i1}(t)$ is always equal to 0 during $I_{i2}(j)$, i.e., $\hat{x}_i(t)= 1$, and $\Delta_{i1}(t)$ is equal to 1 when $x_i(t) = 1$ in $I_{i1}(j)$. We denote the total time duration when $\Delta_{i1}(t)$ is equal to 1 as $T_{e,1}(i,j)$ during the $j$th cycle where $T_{e,1}(i,j) =\sum_{\ell = 1}^{K_1} T_m(i,\ell)$. Thus, we have $\mathbb{E}[T_{e,1}(i)] = \frac{1}{s_i}$. Then, using ergodicity, similar to \cite{bastopcu20e}, $\Delta_{i1}$ is equal to
\begin{align}\label{Delta_i1}
\Delta_{i1}  = \frac{\mathbb{E}[T_{e,1}(i)]}{\mathbb{E}[I_{i}]}= \frac{\mathbb{E}[T_{e,1}(i)]}{\mathbb{E}[I_{i1}]+\mathbb{E}[I_{i2}]}.
\end{align}
Thus, we have 
\begin{align}\label{Delta_i1_val}
\Delta_{i1} = \frac{\mu_i \lambda_i}{\mu_i+\lambda_i}\frac{c_i }{\mu_ic_i+\lambda_i s_i+c_is_i}.
\end{align}

Next, we find $\Delta_{i2}$. We note that $\Delta_{i2}(t)$ is equal to 1 when $x_i(t) = 0$ in $I_{i2}(j)$ and is always equal to 0 during $I_{i1}(j)$. Similarly, we denote the total time duration where $\Delta_{i2}(t)$ is equal to 1 in the $j$th cycle $I_i(j)$ as $T_{e,2}(i,j)$ which is equal to $T_{e,2}(i,j) =\sum_{\ell = 1}^{K_2} T_u(i,\ell)$. Thus, we have $\mathbb{E}[T_{e,2}(i)] = \frac{1}{c_i}$. Then, similar to $\Delta_{i1}$ in (\ref{Delta_i1}), $\Delta_{i2}$ is equal to     
\begin{align}\label{Delta_i2_val}
\Delta_{i2} =\frac{\mu_i \lambda_i}{\mu_i+\lambda_i}\frac{s_i}{\mu_ic_i+\lambda_i s_i+c_is_i}.
\end{align}

By using (\ref{delta_t}), (\ref{Delta_i1_val}), and (\ref{Delta_i2_val}), we obtain $\Delta_i$ as
\begin{align}\label{delta_i}
    \Delta_i = \frac{\mu_i \lambda_i}{\mu_i+\lambda_i}\frac{\theta c_i + (1-\theta)s_i}{\mu_ic_i+\lambda_i s_i+c_is_i}.
\end{align}
Then, by inserting (\ref{delta_i}) in (\ref{total_age}), we obtain $\Delta$. In the next section, we solve the optimization problem in (\ref{problem1}). 

\section{Optimization of Average Difference} \label{sect:opt_soln}
In this section, we solve the optimization problem in (\ref{problem1}). 
Using $\Delta_i$ in (\ref{delta_i}) in (\ref{total_age}), we rewrite (\ref{problem1}) as
\begin{align}
\label{problem1_opt}
\min_{\{s_i, c_i \}}  \quad &  \sum_{i=1}^{n} \frac{\mu_i \lambda_i}{\mu_i+\lambda_i}\frac{\theta c_i + (1-\theta)s_i}{\mu_i c_i+\lambda_i s_i+c_is_i} \nonumber \\
\mbox{s.t.} \quad & \sum_{i=1}^{n} s_i + \sum_{i=1}^{n} c_i\leq C \nonumber \\
\quad & s_i\geq 0, \quad c_i\geq 0,\quad i=1,\dots,n.
\end{align}
We define the Lagrangian function \cite{Boyd04} for (\ref{problem1_opt}) as
\begin{align}\label{lagrange}
    \mathcal{L} =& \sum_{i=1}^{n} \frac{\mu_i \lambda_i}{\mu_i+\lambda_i}\frac{\theta c_i + (1-\theta)s_i}{\mu_i c_i+\lambda_i s_i+c_is_i}+\beta \left(\sum_{i=1}^{n} s_i + c_i- C \right)-\sum_{i=1}^{n}\nu_i s_i -\sum_{i=1}^{n}\eta_i c_i, 
\end{align}
where $\beta \geq 0$, $\nu_i\geq 0$, and $ \eta_i\geq 0$. The KKT conditions are
\begin{align}
\frac{\partial \mathcal{L}}{\partial s_i} =& \frac{\mu_i \lambda_i c_i}{\mu_i+\lambda_i}\frac{(1-\theta)\mu_i -\theta(c_i + \lambda_i)}{(\mu_i c_i+\lambda_i s_i + s_i c_i)^2}+\beta-\nu_i = 0,\label{KKT1}\\
    \frac{\partial \mathcal{L}}{\partial c_i} =& \frac{\mu_i \lambda_i s_i}{\mu_i+\lambda_i}\frac{\theta \lambda_i -(1-\theta)(\mu_i + s_i)}{(\mu_i c_i+\lambda_i s_i + s_i c_i)^2}+\beta-\eta_i = 0,\label{KKT2}
\end{align}
for all $i$. The complementary slackness conditions are 
\begin{align}
    \beta \left(\sum_{i=1}^{n} s_i + c_i- C\right)  = 0,\quad \nu_i s_i = 0, \quad \eta_i c_i =0.\label{CS}
\end{align}

First, we find $s_i$. From (\ref{KKT1}), we have
\begin{align}\label{temp_si}
    (\mu_i c_i+\lambda_i s_i + s_i c_i)^2= \frac{\mu_i \lambda_i c_i}{\mu_i+\lambda_i}\frac{\theta(c_i + \lambda_i)-(1-\theta)\mu_i }{\beta-\nu_i}.
\end{align}
When $\theta(c_i + \lambda_i)\geq (1-\theta)\mu_i$, we solve (\ref{temp_si}) for $s_i$ as 
\begin{align}\label{soln_si}
    s_i = \frac{\mu_i c_i}{\lambda_i+c_i} \left(\!\sqrt{\frac{1}{\mu_i c_i}\frac{\lambda_i}{\mu_i +\lambda_i}\frac{\theta(c_i + \lambda_i)-(1-\theta)\mu_i}{\beta}} -1\!\right)^+\!,
\end{align}
where we used the fact that we either have $s_i>0$ and $\nu_i = 0$, or $s_i = 0$ and $\nu_i\geq 0$, due to (\ref{CS}). Here, $(\cdot)^+ = \max(\cdot,0)$. On the other hand, when $\theta(c_i + \lambda_i)< (1-\theta)\mu_i$, we have $\frac{\partial \Delta_i}{\partial s_i}> 0$, and thus it is optimal to choose $s_i = 0$ as our aim is to minimize $\Delta$ in (\ref{total_age}). In this case, when $s_i = 0$, we have $\Delta_i = \frac{\theta \lambda_i}{\mu_i+\lambda_i}$ which is independent of the value of $c_i$. As we obtain the same $\Delta_i$ for all values of $c_i$, and the total update rate is limited, i.e., $\sum_{i=1}^{n} s_i + c_i \leq  C$, in this case, it is optimal to choose $c_i = 0$ as well (i.e., when $s_i =0$). 

Next, we find $c_i$. From (\ref{KKT2}), we have
\begin{align}\label{ci_eqn}
 (\mu_i c_i+\lambda_i s_i + s_i c_i)^2 = \frac{\mu_i \lambda_i s_i}{\mu_i+\lambda_i}\frac{(1-\theta)(\mu_i + s_i)-\theta \lambda_i }{\beta-\eta_i}.
\end{align}
When $(1-\theta)(\mu_i + s_i)\geq \theta \lambda_i$, we solve (\ref{ci_eqn}) for $c_i$ as 
\begin{align}\label{soln_ci}
c_i = \frac{\lambda_i s_i}{\mu_i + s_i}\left(\!\sqrt{\frac{1}{\lambda_i s_i}\frac{\mu_i}{\mu_i+\lambda_i}\frac{(1-\theta)(s_i+\mu_i)-\theta \lambda_i}{\beta}}-1 \!\right)^+\!,
\end{align}
where we used the fact that we either have $c_i>0$ and $\eta_i = 0$, or $c_i = 0$ and $\eta_i\geq 0$, due to (\ref{CS}). Similarly, when $(1-\theta)(s_i+\mu_i)< \theta \lambda_i$, we have $\frac{\partial \Delta_i}{\partial c_i} >0$. Thus, in this case, it is optimal to choose $c_i = 0$. When $c_i = 0$, we have $\Delta_i = \frac{(1-\theta)\mu_i}{\mu_i+\lambda_i}$ which is independent of the value of $s_i$. Thus, it is optimal to choose $s_i = 0$ when $c_i = 0$.       

From (\ref{soln_si}), if $\frac{1}{\mu_i c_i}\frac{\lambda_i}{\mu_i+\lambda_i}(\theta(c_i + \lambda_i)-(1-\theta)\mu_i)\leq \beta$, we must have $s_i = 0$. Thus, for a given $c_i$, the optimal test rate allocation policy for $s_i$ is a \emph{threshold policy} where $s_i$'s with small $\frac{1}{\mu_i c_i}\frac{\lambda_i}{\mu_i+\lambda_i}(\theta(c_i + \lambda_i)-(1-\theta)\mu_i)$ are equal to zero. Similarly, from (\ref{soln_ci}), if $ \frac{1}{\lambda_i s_i}\frac{\mu_i}{\mu_i+\lambda_i}\left((1-\theta)(s_i+\mu_i)-\theta \lambda_i\right)\leq \beta$, we must have $c_i = 0$. Thus, for a given $s_i$, the optimal policy to determine $c_i$ is a \emph{threshold policy} where $c_i$'s with small $ \frac{1}{\lambda_i s_i}\frac{\mu_i}{\mu_i+\lambda_i}((1-\theta)(s_i+\mu_i)-\theta \lambda_i)$ are equal to zero. 

Next, we show that in the optimal policy, if $s_i>0$ and $c_i>0$ for some $i$, then the total test rate constraint must be satisfied with equality, i.e., $\sum_{i=1}^{n} s_i+c_i = C$.

\begin{lemma}\label{lemma1}
    In the optimal policy, if $s_i>0$ and $c_i>0$ for some $i$, then we have $\sum_{i=1}^{n} s_i+c_i = C$.
\end{lemma}

\begin{Proof}
The derivatives of $\Delta_i$ with respect to $s_i$ and $c_i$ are  
\begin{align}
    \frac{\partial \Delta_i}{\partial s_i} = \frac{\mu_i \lambda_i c_i}{\mu_i + \lambda_i}\frac{(1-\theta)\mu_i-\theta(c_i+\lambda_i)}{\left(c_i \mu_i +s_i c_i+ \lambda_i s_i\right)^2},\\
    \frac{\partial \Delta_i}{\partial c_i} = \frac{\mu_i \lambda_i s_i}{\mu_i + \lambda_i}\frac{\theta\lambda_i-(1-\theta)(s_i+\mu_i)}{\left(c_i \mu_i +s_i c_i+ \lambda_i s_i\right)^2}.
\end{align}
We note that $s_i>0$ in (\ref{soln_si}) implies that $\theta(c_i+\lambda_i) > (1-\theta)\mu_i$. In this case, we have $\frac{\partial \Delta_i}{\partial s_i} <0$. Similarly, $c_i>0$ in (\ref{soln_ci}) implies that $(1-\theta)(s_i+\mu_i)>\theta \lambda_i$. Thus, we have $\frac{\partial \Delta_i}{\partial c_i}<0$. Therefore, in the optimal policy, if we have $s_i>0$ and $c_i>0$ for some $i$, then we must have $\sum_{i=1}^{n} s_i+c_i = C$. Otherwise, we can further decrease $\Delta$ in (\ref{total_age}) by increasing $c_i$ or $s_i$.        
\end{Proof}

Next, we propose an alternating minimization based algorithm for finding $s_i$ and $c_i$. For this purpose, for given initial $(s_i, c_i)$ pairs, we define $\phi_i$ as 
\begin{align} \label{phi-i-defn}
\phi_i \!= \! \begin{cases} 
\frac{1}{\mu_i c_i}\frac{\lambda_i}{\mu_i +\lambda_i}(\theta(c_i + \lambda_i)-(1-\theta)\mu_i), \hspace{1mm} i\!=\!1,\dots,n, \\
\frac{1}{\lambda_i s_i}\frac{\mu_i}{\mu_i+\lambda_i}((1-\theta)(s_i+\mu_i)-\theta \lambda_i), \hspace{1mm} i \!= \!n+1, \dots, 2n.
\end{cases}
\end{align}
Then, we define $u_i$ as
\begin{align}\label{eqn_ui}
 u_i =  \begin{cases} 
\frac{\mu_i c_i}{\lambda_i+c_i}\left( \sqrt{\frac{\phi_i}{\beta}}-1\right)^+, & i=1,\dots,n, \\
\frac{\lambda_i s_i}{\mu_i +s_i}\left(\sqrt{\frac{\phi_i}{\beta}}-1 \right)^+, & i = n+1, \dots, 2n.
\end{cases}   
\end{align}
From (\ref{soln_si}) and (\ref{soln_ci}), $s_i = u_i$ and $c_i = u_{n+i}$, for $i = 1,\dots,n$.

Next, we find $s_i$ and $c_i$ by determining $\beta$ in (\ref{eqn_ui}). First, assume that, in the optimal policy, there is an $i$ such that $s_i>0$ and $c_i>0$. Thus, by Lemma~\ref{lemma1}, we must have $\sum_{i=1}^{n}s_i+c_i = C$. We initially take random $(s_i, c_i)$ pairs such that $\sum_{i=1}^{n}s_i+c_i = C$. Then, given the initial $(s_i, c_i)$ pairs, we immediately choose $u_i =0$ for $\phi_i<0$. For the remaining $u_i$ with $\phi_i\geq0$, we apply a solution method similar to that in \cite{bastopcu20e}. By assuming $\phi_i\geq \beta$, i.e., by disregarding $(\cdot)^+$ in (\ref{eqn_ui}), we solve $\sum_{i=1}^{2n}u_i = C$ for $\beta$. Then, we compare the smallest $\phi_i$ which is larger than zero in (\ref{phi-i-defn}) with $\beta$. If we have $\phi_i \geq \beta$, then it implies that $u_i\geq 0 $ for all remaining $i$. Thus, we have obtained $u_i$ values for given initial ($s_i, c_i$) pairs. If the smallest $\phi_i$ which is larger than zero is smaller than $\beta$, then the corresponding $u_i$ is negative and we should choose $u_i =0$ for the smallest non-negative $\phi_i$. Then, we repeat this procedure until the smallest non-negative $\phi_i$ is larger than $\beta$. After determining all $u_i$, we obtain $s_i = u_i$ and $c_i = u_{n+i} $ for $i = 1,\dots, n$. Then, with the updated values of $(s_i, c_i)$ pairs, we keep finding $u_i$'s until the KKT conditions in (\ref{KKT1}) and (\ref{KKT2}) are satisfied. 

We note that for indices (persons) $i$ for which $(s_i,c_i)$ are zero, the health care provider does not perform any tests, and maps these people as either always infected, i.e., $\hat{x}_i(t) =1$ for all $t$, or always healthy, i.e., $\hat{x}_i(t) =0$. If $\hat{x}_i(t) =0$ for all $t$, $\Delta_i = \frac{\theta \lambda_i}{\mu_i+\lambda_i}$, and if $\hat{x}_i(t) =1$ for all $t$,  $\Delta_i = \frac{(1-\theta) \mu_i}{\mu_i+\lambda_i}$. Thus, for such $i$, the health care provider should choose $\hat{x}_i(t) =0$ for all $t$, if $\frac{\theta \lambda_i}{\mu_i+\lambda_i}< \frac{(1-\theta) \mu_i}{\mu_i+\lambda_i}$, and should choose $\hat{x}_i(t) =1$ for all $t$, otherwise, without performing any tests. 

Finally, we note that the problem in (\ref{problem1_opt}) is not a convex optimization problem as the objective function is not jointly convex in $s_i$ and $c_i$. Therefore, the solutions obtained via the proposed method may not be globally optimal. For that reason, we choose different initial starting points and apply the proposed alternating minimization based algorithm and choose the solution that achieves the smallest $\Delta$ in (\ref{total_age}).  

In the next section, we first provide an alternative way to find the average difference $\Delta$ in (\ref{long_term}) and then characterize the average difference for the erroneous test measurements. 

\section{Average Difference for the Case with Erroneous Test Measurements} \label{sect:error}
We note that the infection status of the $i$th person and its estimate at the health care provider form a continuous time Markov chain \cite[Section 7.5]{bertsekas2008} with the states $(x_i(t), \hat{x}_i(t))\in \{(0,0),\\(0,1),(1,0),(1,1)\}$. In this section, by finding the steady-state distribution for $(x_i(t), \hat{x}_i(t))$, we provide an alternative method to find $\Delta$ in (\ref{long_term}). Then, we consider the case with erroneous test measurements. For this case, we characterize the long-term average difference for the $i$th person denoted by $\Delta_{i}^e$.   

\subsection{An Alternative Method to Characterize Average Difference}
\begin{figure}[t]
\begin{center}
\subfigure[]
{\includegraphics[width=0.48\linewidth]{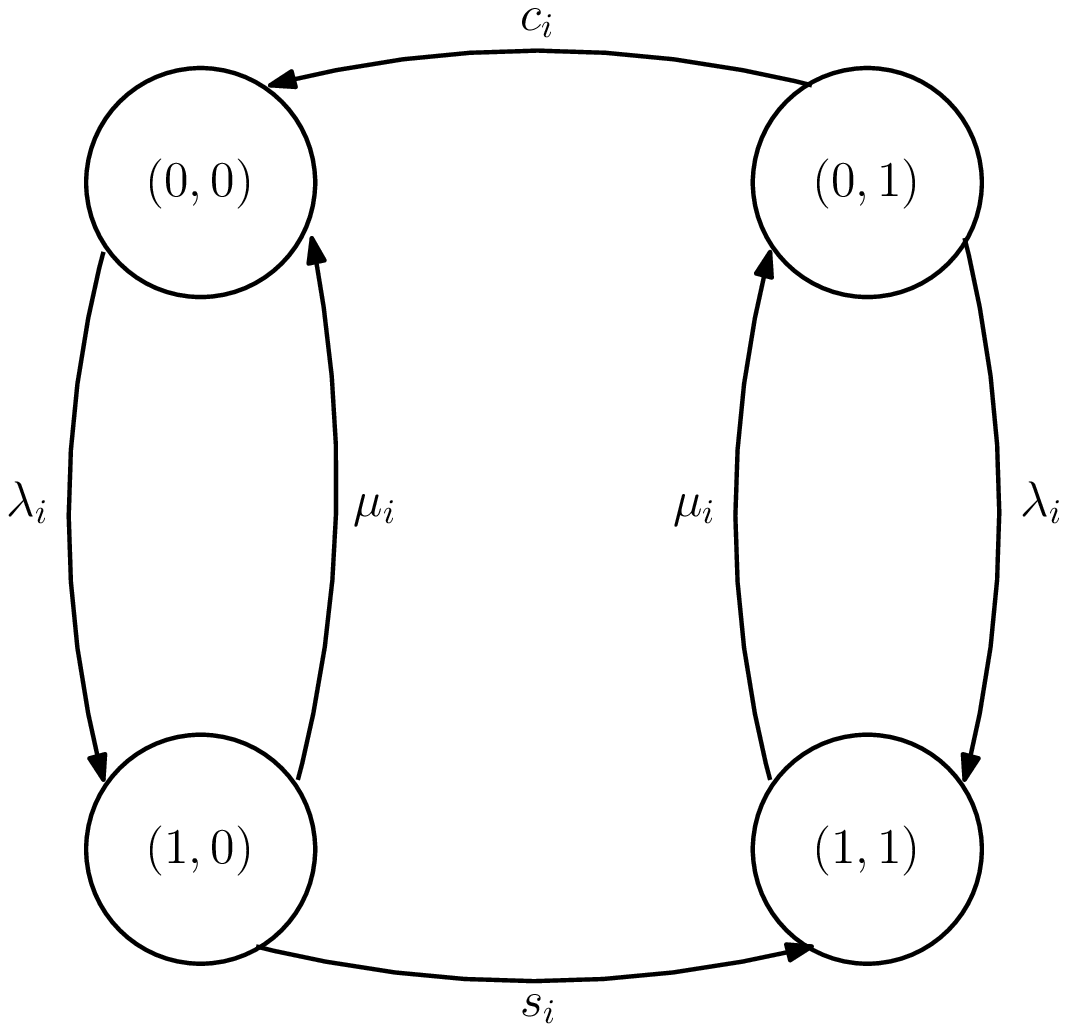}}\hfill 
\subfigure[]
{\includegraphics[width=0.48\linewidth]{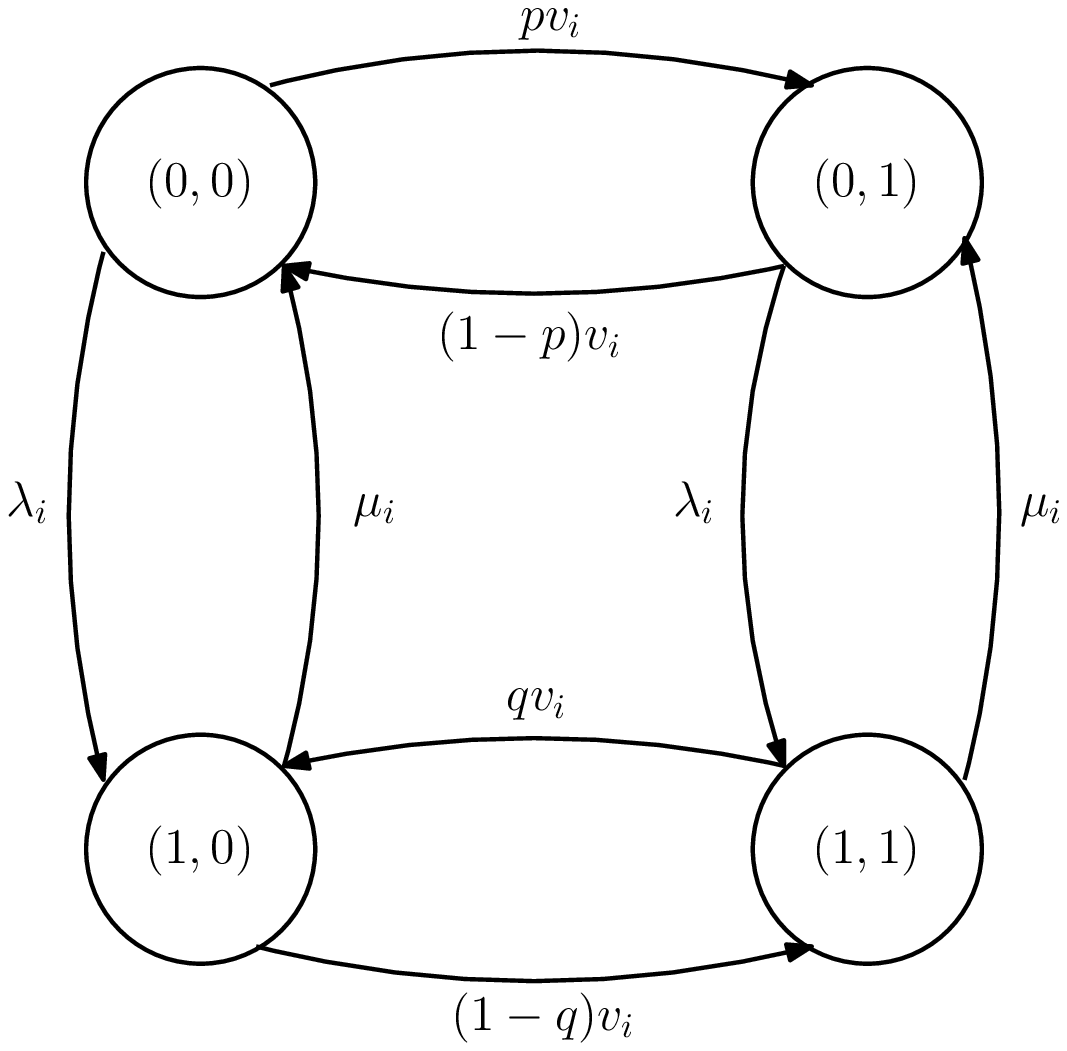}}
\end{center}
\caption{Transition graphs of the states $(x_i(t), \hat{x}_i(t))$ (a) when there is no error in the tests, and (b) when there are errors in the tests.}
\label{fig:Markov}
\end{figure}

When there is no error in the tests, the state transition graph is shown in Fig.~\ref{fig:Markov}(a). Assuming that $s_i> 0 $, $c_i>0$, every state is accessible from any other state, and thus, the Markov chain induced by the system is irreducible. Note that in Section~\ref{sect:opt_soln}, we see that the testing rates for some people can be equal to 0, i.e., $s_i = 0$ and $c_i =0$. For these people, we choose $\hat{x}_i(t)$ to be either always 0 or 1, i.e., consider them as always healthy or sick all the time. Depending on the choice of $\hat{x}_i(t)$, when $s_i = 0$ and $c_i =0$, either the states $(0,0)$ and $(1,0)$, or the states $(0,1)$ and $(1,1)$ will be transient, and thus, have 0 probability in the steady-state. By using small time-step approximation to a discrete time Markov chain, one can show that the self transition probabilities are non-zero, and thus, the Markov chain induced by the system is also aperiodic \cite[Section 7.5]{bertsekas2008}. Therefore, the Markov chain shown in Fig.~\ref{fig:Markov}(a) admits a unique stationary distribution given by $\boldsymbol{\pi} = \{\pi_{00}, \pi_{01}, \pi_{10},\pi_{11}\}$. We find the stationary distribution by writing the local-balance equations which are given as
\begin{align}
    \pi_{00} \lambda_i =& \pi_{10}\mu_i+\pi_{01}c_i, \label{eqn_loc_1}\\
    \pi_{10} (\mu_i+s_i) =& \pi_{00}\lambda_i,\\
    \pi_{01} (c_i+\lambda_i) =& \pi_{11}\mu_i,\\
    \pi_{11} \mu_i =& \pi_{10}s_i +\pi_{01}\lambda_i.\label{eqn_loc_4}
\end{align}

By using (\ref{eqn_loc_1})-(\ref{eqn_loc_4}) and $\sum_{k=1}^{2}\sum_{\ell=1}^{2} \pi_{k\ell} = 1$, we find the steady-state distribution $\boldsymbol{\pi}$ as 
\begin{align}
    \pi_{01} =& \frac{\mu_i\lambda_i}{\mu_i+\lambda_i} \frac{s_i}{\mu_i c_i+\lambda_i s_i+c_is_i },\label{eqn_pi_01}\\
    \pi_{10} =& \frac{\mu_i\lambda_i}{\mu_i+\lambda_i} \frac{c_i}{\mu_i c_i+\lambda_i s_i+c_is_i },\label{eqn_pi_10}
\end{align}
and $\pi_{00}= \frac{\mu_i+s_i}{\lambda_i}\pi_{10}$, and $\pi_{11}=\frac{c_i+\lambda_i}{\mu_i}\pi_{01}$. We note that $\Delta_{i1}$ in (\ref{Delta_i1_val}) is also equal to $\pi_{10}$ in (\ref{eqn_pi_10}), i.e., we have $\Delta_{i1} =\pi_{10}$. Similarly, $\Delta_{i2}$ in (\ref{Delta_i2_val}) is equal to $\pi_{01}$ in (\ref{eqn_pi_01}). Thus, by observing that the states $(x_i(t),\hat{x}_i(t))$ form a continuous time Markov chain, we can find the average difference $\Delta$ in (\ref{long_term}) by finding the steady-state distribution for $\boldsymbol{\pi}$. This method will be particularly useful in the following section where we consider the case with erroneous test measurements.

\subsection{Average Difference with Erroneous Test Measurements}
In this section, we consider the case where the test measurements can be erroneous. When a test in applied to an infected person, i.e., when $x_i(t) = 1$, the test result will be 0 with probability $q$ and 1 with probability $1-q$, where $0 \leq q<\frac{1}{2}$. In other words, the false negative probability is equal to $q$. Similarly, when a test is applied to a healthy person, i.e., when $x_i(t) = 0$, the test result will be $1$ with probability $p$ and $0$ with probability $1-p$, where $0\leq p<\frac{1}{2}$. Thus, the false positive probability is equal to $p$. The probability distribution for the test measurements is provided in Table~\ref{table:tests}.

\begin{table}[h]
\small
\centering
	\begin{center}
		\begin{tabular}{ | c | c | c |}
			\hline
			$x_i(t)$ $\setminus$ $\hat{x}_i(t)$ & $ 0$ & $1$ \\ \hline
			$0$  & $ 1-p$ & $p$ \\ \hline
			$1$ & $ q$ & $1-q$ \\ \hline
		\end{tabular}
	\end{center}
	\caption{The probability distribution for successful and false test measurements.}
	\label{table:tests}
	\vspace{-0.4cm}
\end{table}

In this section, we consider the case where the health care provider applies only one test rate $v_i$ to the $i$th person, whether the person is currently marked as healthy or infected. That is, we do not consider separate testing rates of $s_i$ and $c_i$ for healthy and infected people as we did before, instead, here both $s_i$ and $c_i$ are equal o $v_i$. Since the health care provider applies the same test rate for the $i$th person, here we do not consider the importance factor $\theta$ either. Then, we define the long-term average difference for the $i$th person with the error on the test measurements as follows, where the superscript $e$ stands for ``erroneous''
\begin{align}\label{Delta_ei_defn}
    \Delta_{i}^e = \Delta_{i1}^e + \Delta_{i2}^e,
\end{align}
and the definitions of $\Delta_{i1}^e$ and $\Delta_{i2}^e$ follow similarly from (\ref{Delta_i1}). We note that with the test rates $v_i$ and errors on the test measurements, the states $(x_i(t), \hat{x}_i(t))$ form a continuous time Markov chain, and the corresponding state transition graph is shown in Fig.~\ref{fig:Markov}(b). Assuming that $v_i>0$, one can show that there is a unique steady-state distribution $\boldsymbol{\pi^e} = \{\pi^e_{00},\pi^e_{01},\pi^e_{10},\pi^e_{11}\}$ which can be found by solving the local balance equations which are given as follows
\begin{align}
    \pi^e_{00} (v_ip +\lambda_i) =& \pi^e_{01} v_i(1-p)+\pi^e_{10} \mu_i,\label{eqn_loc2_1} \\ 
    \pi^e_{10} (v_i(1-q) +\mu_i) =& \pi^e_{00} \lambda_i+\pi^e_{11} v_i q, \\
    \pi^e_{01} (v_i(1-p) +\lambda_i) =& \pi^e_{00} v_i p+\pi^e_{11} \mu_i, \\
    \pi^e_{11} (v_i q +\mu_i) =& \pi^e_{10} v_i(1-q)+\pi^e_{01} \lambda_i . \label{eqn_loc2_4}
\end{align}
Then, by using (\ref{eqn_loc2_1})-(\ref{eqn_loc2_4}) and $\sum_{k=1}^{2}\sum_{\ell=1}^{2} \pi^e_{k\ell} = 1$, we find the steady-state distribution $\boldsymbol{\pi^e}$ as
\begin{align}
    \pi^e_{00} =& \frac{\mu_i\lambda_i q+(1-p)\mu_i(v_i+\mu_i)}{(\lambda_i+\mu_i)(\lambda_i+\mu_i+v_i)}, \\ 
    \pi^e_{01} =& \frac{\mu_i\lambda_i (1-q)+p\mu_i(v_i+\mu_i)}{(\lambda_i+\mu_i)(\lambda_i+\mu_i+v_i)}, \label{eqn_pi_e_01} \\
     \pi^e_{10} =& \frac{\mu_i\lambda_i (1-p)+q\lambda_i(v_i+\lambda_i)}{(\lambda_i+\mu_i)(\lambda_i+\mu_i+v_i)},\label{eqn_pi_e_10} \\
    \pi^e_{11} =& \frac{\mu_i\lambda_i p+(1-q)\lambda_i(v_i+\lambda_i)}{(\lambda_i+\mu_i)(\lambda_i+\mu_i+v_i)}.
\end{align}

We note that $\Delta_{i1}^e$, and $\Delta_{i2}^e$ are equal to $\pi^e_{10}$ in (\ref{eqn_pi_e_10}), and $\pi^e_{01}$ in (\ref{eqn_pi_e_01}), respectively. Thus, if $v_i>0$, then $\Delta_{i}^e$ in (\ref{Delta_ei_defn}) becomes 
\begin{align}\label{Delta_ei_val}
    \Delta_{i}^e = \frac{p \mu_i^2 + q\lambda_i^2+(2-p-q)\mu_i\lambda_i +v_i(p \mu_i + q\lambda_i) }{(\lambda_i+\mu_i)(\lambda_i+\mu_i+v_i)}.
\end{align}

We immediately note that if false positive test probability $p$ and false negative test probability $q$ are equal to 0, $\Delta_{i}^e$ becomes $\frac{2\mu_i\lambda_i }{(\lambda_i+\mu_i)(\lambda_i+\mu_i+v_i)}$ which is equal to $\Delta_{i1} + \Delta_{i2}$ provided in (\ref{Delta_i1_val}) and (\ref{Delta_i2_val}), respectively when $v_i = s_i = c_i$. Then, $\frac{\partial \Delta_{i}^e}{\partial p}\geq 0$ is equivalent to $v_i+\mu_i -\lambda_i \geq 0$ and $\frac{\partial \Delta_{i}^e}{\partial q}\geq 0$ is equivalent to $v_i+\lambda_i-\mu_i \geq 0$ which means that depending on the values of $v_i$, $\mu_i$, and $\lambda_i$, the long-term average difference $\Delta_{i}^e$ can be an increasing function of only $p$ or only $q$, or both $p$ and $q$, but $\Delta_{i}^e$ cannot be a decreasing function of both $p$ and $q$. This is expected as false negative and false positive tests negatively affect the estimation process. One can also show that $\frac{\partial \Delta_{i}^e}{\partial v_i}<0$ and $\frac{\partial^2 \Delta_{i}^e}{\partial v_i^2}>0$ which means that $\Delta_{i}^e$ decreases with $v_i $ and is a convex function of the test rate $v_i$. 

Next, we consider the case when $v_i = 0$. Note that when $v_i = 0$, the health care provider either maps these people as always sick or always healthy depending on their infection and recovery rates. Thus, when $v_i = 0$ and depending on the estimate $\hat{x}_i(t)$, two of the states in Fig.~\ref{fig:Markov}(b) will never be visited and thus, these states will have 0 steady-state probabilities. For this case, the steady states are given by $\bar{\pi}^e_{1,\hat{x}_i}$ and $\bar{\pi}^e_{0,\hat{x}_i}$. The local balance equation is $\lambda_i \bar{\pi}^e_{0,\hat{x}_i} = \mu_i \bar{\pi}^e_{1,\hat{x}_i}$. By using $\bar{\pi}^e_{0,\hat{x}_i}+\bar{\pi}^e_{1,\hat{x}_i}=1$, we find the steady-state distribution as $\bar{\pi}^e_{0,\hat{x}_i} = \frac{\mu_i}{\mu_i+\lambda_i}$, and $\bar{\pi}^e_{1,\hat{x}_i} = \frac{\lambda_i}{\mu_i+\lambda_i}$. Thus, if $\mu_i <\lambda_i$, i.e., if people are infected more frequently, then the health care provider chooses its estimate as $\hat{x}_i(t) = 1$ and, $\Delta_{i}^e =  \frac{\mu_i}{\mu_i+\lambda_i}$. If $\mu_i \geq\lambda_i$, i.e., if people stay healthy more often, then we have $\hat{x}_i(t) = 0$, and $\Delta_{i}^e =  \frac{\lambda_i}{\mu_i+\lambda_i}$. Therefore, when $v_i =0$, we have
\begin{align} \label{Delta_ei_val_alt}
  \Delta_{i}^e =\min \left\{\frac{\mu_i}{\mu_i+\lambda_i}, \frac{\lambda_i}{\mu_i+\lambda_i}\right\}.  
\end{align}

In order to find the optimal test rates $v_i$ in the case of errors on the test measurements, we formulate the following optimization problem      
\begin{align}
\label{problem2_opt}
\min_{\{v_i \}}  \quad &  \sum_{i=1}^{n}\mathbbm{1}\{v_i>0\} \frac{p \mu_i^2 + q\lambda_i^2+(2-p-q)\mu_i\lambda_i +v_i(p \mu_i + q\lambda_i) }{(\lambda_i+\mu_i)(\lambda_i+\mu_i+v_i)} \nonumber\\ 
&\qquad +\mathbbm{1}\{v_i=0\}\min \left\{\frac{\mu_i}{\mu_i+\lambda_i}, \frac{\lambda_i}{\mu_i+\lambda_i} \right\} \nonumber \\
\mbox{s.t.} \quad & \sum_{i=1}^{n} v_i\leq C \nonumber \\
\quad & v_i\geq 0,\quad i=1,\dots,n,
\end{align}
where the objective function is given by the summation of $\Delta_{i}^e$ in (\ref{Delta_ei_val}) when $v_i>0$ and $\Delta_{i}^e$ in (\ref{Delta_ei_val_alt}) when $v_i=0$ over all people and $\mathbbm{1}\{.\}$ is the indicator function taking value 1 when $\{\cdot\}$ is true and 0, otherwise. In (\ref{problem2_opt}), we have a constraint on the total test rate, i.e., $\sum_{i=1}^{n} v_i\leq C $. We note that the optimization problem in (\ref{problem2_opt}) is in general not convex due to the indicator function in the objective function. However, for a given set of $\mathbbm{1}\{v_i=0\}$, the optimization problem in (\ref{problem2_opt}) is convex and can be solved optimally. Thus, by solving the problem in (\ref{problem2_opt}) for all possible set of  $\mathbbm{1}\{v_i=0\}$, we can determine the global optimal solution which requires to solve $2^n$ different optimization problems which can be impractical for large $n$. Because of this reason, next, we provide a greedy algorithm to solve the optimization problem in (\ref{problem2_opt}).    

In the greedy solution, initially, assuming that $\mathbbm{1}\{v_i>0\} =1$ for all $i$, we consider the following the optimization problem
\begin{align}
\label{problem2_opt_alt1}
\min_{\{v_i \}}  \quad &  \sum_{i=1}^{n} \frac{p \mu_i^2 + q\lambda_i^2+(2-p-q)\mu_i\lambda_i +v_i(p \mu_i + q\lambda_i) }{(\lambda_i+\mu_i)(\lambda_i+\mu_i+v_i)} \nonumber \\
\mbox{s.t.} \quad & \sum_{i=1}^{n} v_i\leq C \nonumber \\
\quad & v_i\geq 0,\quad i=1,\dots,n,
\end{align}
where the objective function in (\ref{problem2_opt_alt1}) is equal to $\Delta_{i}^e$ in (\ref{Delta_ei_val}). For this optimization problem, we define the Lagrangian function for (\ref{problem2_opt_alt1}) as
\begin{align}\label{lagrange2}
    \mathcal{L} =& \sum_{i=1}^{n} \frac{p \mu_i^2 + q\lambda_i^2+(2-p-q)\mu_i\lambda_i +v_i(p \mu_i + q\lambda_i) }{(\lambda_i+\mu_i)(\lambda_i+\mu_i+v_i)}+\bar{\beta} \left(\sum_{i=1}^{n}v_i- C \right)-\sum_{i=1}^{n}\bar{\nu}_i v_i, 
\end{align}
where $\bar{\beta} \geq 0$, $\bar{\nu}_i\geq 0$. We note that the problem defined in (\ref{problem2_opt_alt1}) is a convex optimization problem, and thus we can find the optimal test rates $v_i$ by analyzing the KKT and the complementary slackness conditions. The KKT conditions are given by 
\begin{align}
\frac{\partial \mathcal{L}}{\partial v_i} =& \frac{-2(1-p-q)\mu_i \lambda_i }{(\mu_i+\lambda_i)(\mu_i+\lambda_i+v_i)^2}+\bar{\beta}-\bar{\nu}_i = 0,\label{KKT1_problem2}
\end{align}
for all $i$. The complementary slackness conditions are 
\begin{align}
   \bar{\beta} \left(\sum_{i=1}^{n} v_i- C\right)  = 0,\quad \bar{\nu}_i v_i = 0.\label{CS_problem2}
\end{align}
By using (\ref{KKT1_problem2}) and (\ref{CS_problem2}), we find the optimal $v_i $ values for the problem in (\ref{problem2_opt_alt1}) as   
\begin{align}\label{soln_si_bar}
    v_i = (\mu_i +\lambda_i ) \left(\sqrt{\frac{\mu_i\lambda_i}{(\mu_i+\lambda_i)^3}\frac{2(1-p-q)}{\bar{\beta} }} -1\right)^+.
\end{align}

With the test rates $v_i $ in (\ref{soln_si_bar}) we find the average differences $\Delta_{i}^e$ in (\ref{Delta_ei_val}) and then compare them with $\Delta_{i}^e$ in (\ref{Delta_ei_val_alt}) when $v_i = 0$. Due to the errors in the tests, $\Delta_{i}^e$ in (\ref{Delta_ei_val_alt}) with $v_i = 0$ can be smaller than $\Delta_{i}^e$ in (\ref{Delta_ei_val}) with the test rates $v_i $ found in (\ref{soln_si_bar}). For these people, we choose index $i$ where the difference between $\Delta_{i}^e$ in (\ref{Delta_ei_val}) with the $v_i $ in (\ref{soln_si_bar}) and $\Delta_{i}^e$ in (\ref{Delta_ei_val_alt}) is the highest. Then, we take $v_i= 0$ as applying no test to this person can further decrease $\Delta_{i}^e$. For the remaining people, we solve the optimization problem in (\ref{problem2_opt_alt1}). After obtaining the test rates for the remaining people, we again compare average differences $\Delta_{i}^e$ with the test rates in (\ref{soln_si_bar}) and with no test and we choose $v_i= 0$ for the person where $\Delta_{i}^e$ can be further decreased. We repeat these steps until all $\Delta_{i}^e$s with $v_i> 0$ cannot be further decreased by choosing $v_i= 0$.                

We note that the solution obtained in (\ref{soln_si_bar}) has a \emph{threshold} structure. As false positive and negative test rates increase, the term $\frac{2(1-p-q)}{\bar{\beta} }$ in (\ref{soln_si_bar}) gets smaller. As a result, some people with higher $\sqrt{\frac{(\mu_i+\lambda_i)^3}{\mu_i\lambda_i}}$  may not be tested by the health care provider. Thus, when $p$ and $q$ are high, a smaller portion of the population is tested with higher test rates in order to combat the test errors. 

\section{Average Estimation Error with Dependent Infection Rates}\label{sect:Dependent}
\begin{figure}[t]
	\centerline{\includegraphics[width=0.75\columnwidth]{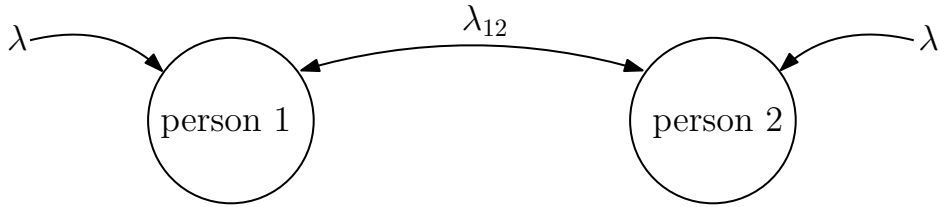}}
	\caption{The infection rates of two people where the individual infection rate is equal to $\lambda$. When the infection has not been detected, these two people can infect each other with rate $\lambda_{12}$.}
	\label{Fig:dependent_inf_fig}
\end{figure}

In this section, we consider the case where we have two people whose infection rates depend on each other. When these two people are healthy, they can be individually infected with the virus after an exponential time with rate $\lambda$. When one of these two people is infected and this has not been detected by the health care provider, this person can infect the other healthy person after an exponential time with rate $\lambda_{12}$ which has been illustrated in Fig.~\ref{Fig:dependent_inf_fig}. Thus, when both of the people are healthy, their individual infection rate is $\lambda$. However, when one of them is sick and this has not been detected by the health care provider, the healthy person's total infection rate is equal to $\lambda+\lambda_{12}$. On the other hand, if only one person is infected, i.e., $x_i(t) = 1$, which has also been detected by the health care provider, $\hat{x}_i(t) = 1$, then we assume that we isolate the infected person from the healthy one, and thus, the healthy person's infection rate remains as $\lambda$ instead of $\lambda+\lambda_{12}$. When the people are infected, they recover from the disease after an exponential time with rate $\mu$. 

When the health care provider thinks that a person is healthy, i.e., $\hat{x}_i(t) = 0$, the next test is applied to this person after an exponential time with rate $s$. When the health care provider thinks that a person is sick, i.e., $\hat{x}_i(t) = 1$, the next test applied to this person after an exponential time with rate $c$. Here, we note that since the people are identical in terms of their infection and the recovery rates, the health care provider applies the same test rates.

Similar to Section~\ref{sect:error}, we note that the states $\{x_1(t),\hat{x}_1(t),x_2(t),\hat{x}_2(t)\}$ form a continuous time Markov chain where the unique stationary distribution is given by $\boldsymbol{\pi^d} = \{\pi_{0000}^d,\pi_{0001}^d, \dots, \pi_{1111}^d\}$. In order to find the stationary distribution, we write the local balance equations as follows
\begin{align}
  2\lambda \pi_{0000}^d  =& \mu \pi_{1000}^d + c\pi_{0100}^d + \mu \pi_{0010}^d+c\pi_{0001}^d, \label{eqn_dep_1} \\
  (2\lambda + c) \pi_{0001}^d =& \mu \pi_{0011}^d+ c \pi_{0101}^d + \mu \pi_{1001}^d,\\
  (\lambda+\lambda_{12}+\mu+s)\pi_{0010}^d =& c \pi_{0110}^d+ \mu \pi_{1010}^d + \lambda \pi_{0000}^d,\\
 (\lambda+ \mu) \pi_{0011}^d =& c\pi_{0111}^d + \mu \pi_{1011}^d + s\pi_{0010}^d+ \lambda\pi_{0001}^d,\\ 
 (2\lambda + c) \pi_{0100}^d =& c\pi_{0101}^d+ \mu \pi_{0110}^d + \mu \pi_{1100}^d,\\ 
 (2\lambda + 2c)\pi_{0101}^d =& \mu \pi_{0111}^d + \mu \pi_{1101}^d,\\
 (\lambda + \mu +s+ c) \pi_{0110}^d =& \lambda \pi_{0100}^d+\mu \pi_{1110}^d, \\ 
 (\lambda +\mu+c) \pi_{0111}^d =& s\pi_{0110}^d + \lambda \pi_{0101}^d + \mu \pi_{1111}^d,\\
 (\lambda +\lambda_{12}+\mu+s) \pi_{1000}^d=& \lambda \pi_{0000}^d + c\pi_{1001}^d + \mu \pi_{1010}^d, \\ 
 (\lambda+\mu+s+c) \pi_{1001}^d = & \mu \pi_{1011}^d + \lambda \pi_{0001}^d,\\ 
 (2\mu + 2s) \pi_{1010}^d =& (\lambda + \lambda_{12})\pi_{1000}^d + (\lambda + \lambda_{12})\pi_{0010}^d,\\
 (2\mu + s)\pi_{1011}^d =& s\pi_{1010}^d + \lambda \pi_{1001}^d + \lambda \pi_{0011}^d, \\ 
 ( \lambda + \mu) \pi_{1100}^d =& s \pi_{1000}^d + \lambda\pi_{0100}^d + c \pi_{1101}^d + \mu \pi_{1110}^d,\\
 (\lambda + \mu + c) \pi_{1101}^d =& s \pi_{1001}^d + \lambda \pi_{0101}^d + \mu \pi_{1111}^d, \\ 
 (2\mu +s) \pi_{1110}^d =& \lambda \pi_{1100}^d + s \pi_{1010}^d + \lambda \pi_{0110}^d, \\ 
 2\mu \pi_{1111}^d =& s\pi_{1110}^d + \lambda \pi_{1101}^d + s\pi_{1011}^d + \lambda \pi_{0111}^d.\label{eqn_dep_16} 
\end{align}

By using (\ref{eqn_dep_1})-(\ref{eqn_dep_16}) and $\sum_{j=1}^{2}\sum_{\ell=1}^{2}\sum_{m=1}^{2}\sum_{h=1}^{2}\pi_{j \ell m h}^d = 1$, we find the stationary distribution $\boldsymbol{\pi^d}$. We denote the long-term average estimation error for person $i$ as $\Delta_{i}^d$ for $i=1,2$, where the superscript $d$ stands for ``dependent'', which is given by 
\begin{align}
    \Delta_{i}^d = \Delta_{i1}^d+\Delta_{i2}^d, 
\end{align}
where $\Delta_{i1}^d$ and $\Delta_{i2}^d$ follow from (\ref{Delta_i1}). Then, we have 
\begin{align}
    \Delta_{11}^d = &\pi_{1000}^d + \pi_{1001}^d+\pi_{1010}^d+ \pi_{1011}^d, \\
    \Delta_{12}^d = &\pi_{0100}^d + \pi_{0101}^d+\pi_{0110}^d+ \pi_{0111}^d, \\ 
    \Delta_{21}^d = &\pi_{0010}^d + \pi_{0110}^d+\pi_{1010}^d+ \pi_{1110}^d, \\
    \Delta_{22}^d = &\pi_{0001}^d + \pi_{0101}^d+\pi_{1001}^d+ \pi_{1101}^d.
\end{align}
In Section~\ref{sect:num_res}, for given infection, recovery and test rates, we numerically evaluate the stationary distribution and find the average difference $\Delta_{i}^d$. 

\section{Age of Incorrect Information Based Error Metric}\label{sect:aoii_model}
So far, we have considered an estimation error metric that takes value 1 if the actual infection status of a person is different than the real-time estimation at the health care provider. Thus, the error metric takes values based on the information content. On the other hand, the traditional age metric introduced in \cite{Kaul12a} considers only the time passed since the most recently received status update packet is generated at the source. As a result, the traditional age metric does not consider the information content and age alone may not be a suitable performance metric for the problem considered in our work.

\begin{figure}[t]
\begin{center}
\subfigure[]
{\includegraphics[width=0.49\linewidth]{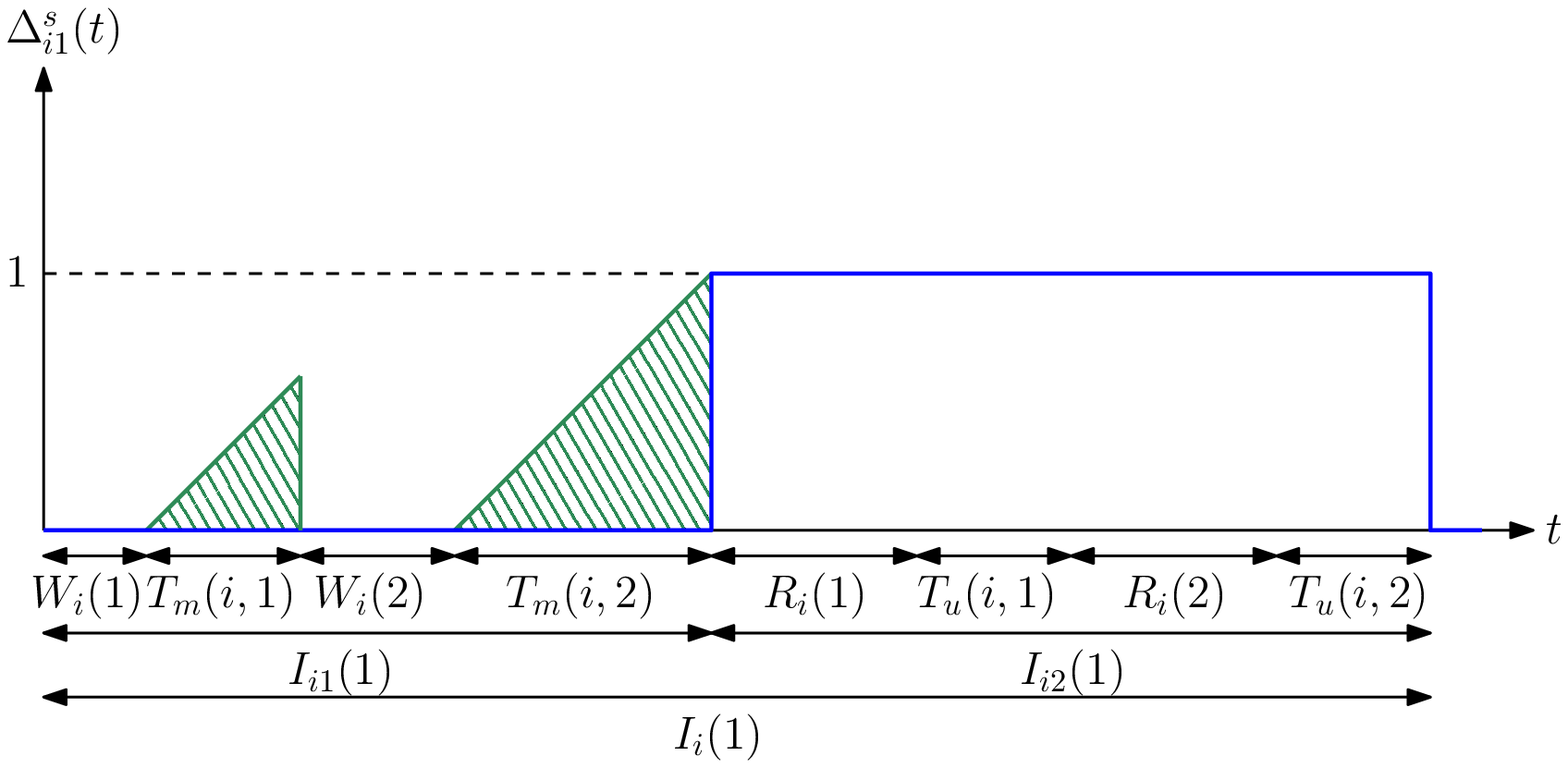}} 
\subfigure[]
{\includegraphics[width=0.49\linewidth]{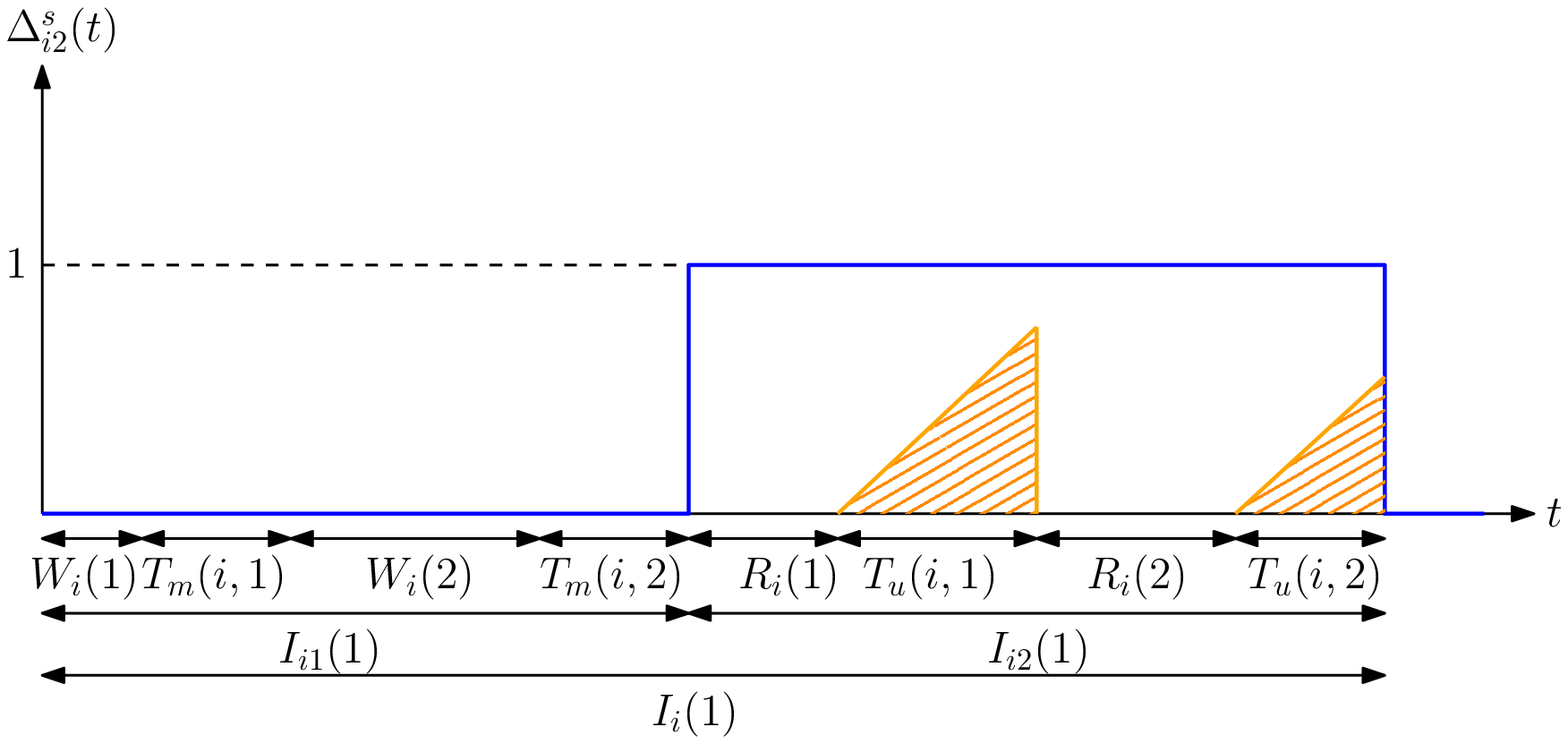}}
\end{center}
\caption{A sample evolution of (a) $\Delta_{i1}^s(t)$, and (b) $\Delta_{i2}^s(t)$ in a typical update cycle.}
\label{fig:error_aoii}
\end{figure}

In the context of infection tracking, it is important to know how long the estimation at the health care provider has been different from the actual infection status of the people. However, the error metric that we have considered so far does not have the time component, i.e., it only takes value 1 independent of the time duration that it has been off from the actual health status. Motivated by the AoII introduced in \cite{Maatouk20a, Maatouk20b} which takes into account both the time and the information content, in this section, we consider the following error metric, where the superscript $s$ stands for ``synchronization'' implied in AoII,
\begin{align}\label{eqn:AoII}
    \Delta_{i}^s = (t-V_i(t))\mathbbm{1}\{\hat{x}_i(t)\neq x_i(t)\}, 
\end{align}
where $V_i(t)$ is the last time instant where the health care provider has the accurate estimation of the health status for the $i$th person, i.e., the last time instant when $\Delta_{i}^s = 0$. Similarly, we define 
\begin{align}
    \Delta_{i1}^s =& (t-V_{i1}(t))\max\{x_i(t) - \hat{x}_i(t),0\}, \label{eqn:AoII_1}\\
    \Delta_{i2}^s =& (t-V_{i2}(t))\max\{\hat{x}_i(t) - x_i(t),0\}, \label{eqn:AoII_2}
\end{align}
where $V_{i1}(t)$ and $V_{i2}(t)$ are equal to the last time instants when $\Delta_{i1}^s$ and $\Delta_{i2}^s$ are equal to 0, respectively. A sample evolution of $\Delta_{i1}^s$ and $\Delta_{i2}^s$ is shown in Fig.~\ref{fig:error_aoii} and we note that $\Delta_{i}^s(t) = \Delta_{i1}^s(t)  + \Delta_{i2}^s(t) $.  

Similar to Section~\ref{Sec:Average_difference}, the infection and the recovery rates of the $i$th person are $\lambda_i$ and $\mu_i$, respectively. In this section, the health care provider applies only one test rate for each person denoted by $w_i$. That is, we do not consider separate testing rates of $s_i$ and $c_i$ for healthy and infected people as we did before, instead, here both $s_i$ and $c_i$ are equal o $w_i$. We first consider the case where $w_i>0$. By following the steps in Section~\ref{Sec:Average_difference}, one can show that $\mathbb{E}[I_{i1}] = \frac{1}{w_i} + \frac{w_i+\mu_i}{w_i\lambda_i}$ and $\mathbb{E}[I_{i2}] = \frac{1}{w_i} + \frac{w_i+\lambda_i}{w_i\mu_i}$ which can be obtained by substituting $w_i$ instead of $s_i$ and $c_i$ in (\ref{I_1}) and (\ref{I_2}), respectively. Next, we denote the total area when $\Delta_{i1}^s(t)>0$ as $A_{e,1}(i,j)$ during the $j$th cycle where $A_{e,1}(i,j) =\sum_{\ell = 1}^{K_1} \frac{T_m(i,\ell)^2}{2}$ and $K_1$ has a geometric distribution with success rate $\frac{w_i}{\mu_i+w_i}$. Then, we have $\mathbb{E}[A_{e,1}(i)] = \frac{1}{w_i(w_i+\mu_i)}$. Similarly, we denote the total area when $\Delta_{i2}^s(t)>0$ as $A_{e,2}(i,j)$ during the $j$th cycle where $A_{e,2}(i,j) =\sum_{\ell = 1}^{K_2} \frac{T_u(i,\ell)^2}{2}$ and $K_2$ has a geometric distribution with success rate $\frac{w_i}{\lambda_i+w_i}$. Then, we have $\mathbb{E}[A_{e,2}(i)] = \frac{1}{w_i(w_i+\lambda_i)}$. By using ergodicity, the long-term average differences become $\Delta_{i1}^s = \frac{\mathbb{E}[A_{e,1}(i)]}{\mathbb{E}[I_{i1}]+\mathbb{E}[I_{i2}]}$ and $\Delta_{i2}^s = \frac{\mathbb{E}[A_{e,2}(i)]}{\mathbb{E}[I_{i1}]+\mathbb{E}[I_{i2}]}$ which gives 
\begin{align}\label{eqn_aoii_final}
    \Delta_{i}^s = \Delta_{i1}^s+\Delta_{i2}^s = \frac{\mu_i\lambda_i}{\mu_i+\lambda_i}\frac{2w_i+\mu_i+\lambda_i }{(w_i+\mu_i+\lambda_i)(w_i+\mu_i)(w_i+\lambda_i)},
\end{align}
when $w_i>0$. One can show that $\Delta_{i}^s$ is a decreasing function of $w_i$, i.e., $\frac{\partial\Delta_{i}^s}{\partial w_i}<0$, and $\Delta_{i}^s$ is a convex function of $w_i$, i.e., $\frac{\partial^2\Delta_{i}^s}{\partial w_i^2}>0$. 

When $w_i=0$, we have $\mathbb{E}[I_{i}] = \frac{\mu_i \lambda_i}{\mu_i+\lambda_i}$, i.e., $\mathbb{E}[I_{i}]$ is equal to the expected time of a person's healthy and sick states. Since the health care provider applies no tests to test a person, it either estimates this person to be always sick ($\hat{x}_i(t) = 1$) or always healthy ($\hat{x}_i(t) = 0$). When $w_i=0$ and $\hat{x}_i(t) = 1$, then $\Delta_{i}^s = \frac{1}{\mu_i}\frac{\lambda_i}{\mu_i+\lambda_i}$. When $w_i=0$ and $\hat{x}_i(t) = 1$, we have $\Delta_{i}^s = \frac{1}{\lambda_i}\frac{\mu_i}{\mu_i+\lambda_i}$. If $\mu_i<\lambda_i $, then the health care provider $\hat{x}_i(t) = 1$, and $\hat{x}_i(t) = 0$, otherwise. Thus, when $w_i=0$, we have $\Delta_{i}^s = \min\left\{\frac{1}{\mu_i}\frac{\lambda_i}{\mu_i+\lambda_i}, \frac{1}{\lambda_i}\frac{\mu_i}{\mu_i+\lambda_i} \right\}$.   

In order to find the optimal test rates, we formulate the following optimization problem
\begin{align}
\label{problem3_opt}
\min_{\{w_i \}}  \quad &  \sum_{i=1}^{n}\mathbbm{1}\{w_i>0\} \frac{\mu_i\lambda_i}{\mu_i+\lambda_i}\frac{2w_i+\mu_i+\lambda_i }{(w_i+\mu_i+\lambda_i)(w_i+\mu_i)(w_i+\lambda_i)} \nonumber\\ 
&\qquad+\mathbbm{1}\{w_i = 0\}\min\left\{\frac{1}{\mu_i}\frac{\lambda_i}{\mu_i+\lambda_i}, \frac{1}{\lambda_i}\frac{\mu_i}{\mu_i+\lambda_i} \right\} \nonumber \\
\mbox{s.t.} \quad & \sum_{i=1}^{n} w_i\leq C \nonumber \\
\quad & w_i\geq 0,\quad i=1,\dots,n,
\end{align} 
where the objective function in (\ref{problem3_opt}) is equal to the summation of $\Delta_{i}^s$ in (\ref{eqn_aoii_final}) when $w_i>0$ and $\Delta_{i}^s$ when $w_i=0$ over all people. In order to solve the problem in (\ref{problem3_opt}), we follow the same greedy solution approach in Section~\ref{sect:error}. First, by assuming that $w_i>0$, and thus, the average difference $\Delta_{i}^s$ is given in (\ref{eqn_aoii_final}), we solve the following optimization problem 
 
\begin{align}
\label{problem3_opt_alt}
\min_{\{w_i \}}  \quad &  \sum_{i=1}^{n} \frac{\mu_i\lambda_i}{\mu_i+\lambda_i}\frac{2w_i+\mu_i+\lambda_i }{(w_i+\mu_i+\lambda_i)(w_i+\mu_i)(w_i+\lambda_i)} \nonumber \\
\mbox{s.t.} \quad & \sum_{i=1}^{n} w_i\leq C \nonumber \\
\quad & w_i\geq 0,\quad i=1,\dots,n.
\end{align}
Since the problem in (\ref{problem3_opt_alt}) is a convex optimization problem, by defining Lagrangian function and analyzing the KKT and the complementary slackness conditions, we can find the optimal $w_i$ values. In order not to be repetitive, we skip these optimization steps. Then, we compare $\Delta_{i}^s$ in (\ref{eqn_aoii_final}) with $w_i$ values found in (\ref{problem3_opt_alt}) with $\min\{\frac{1}{\mu_i}\frac{\lambda_i}{\mu_i+\lambda_i}, \frac{1}{\lambda_i}\frac{\mu_i}{\mu_i+\lambda_i} \}$. If we can reduce $\Delta_{i}^s$ further, we choose $w_i = 0$ for the person with the highest improvement. Then, we solve the optimization problem in (\ref{problem3_opt_alt}) for the remaining people. We repeat these steps until there is no improvement in $\Delta_{i}^s$ by choosing $w_i =0$.     

In the next section, we provide extensive numerical results to evaluate optimal test rates in various settings considered in this paper.   

\section{Numerical Results} \label{sect:num_res}
In this section, we provide seven numerical results. For these examples, we take $\lambda_i$ as
\begin{align}\label{lambda_i}
    \lambda_i = a r^i, \quad i=1,\dots,n,
\end{align}
where $r= 0.9$ and $a$ is such that $\sum_{i=1}^{n}\lambda_i = 6$. Also, we take $\mu_i$ as
\begin{align}\label{alpha_i}
    \mu_i = b q^i, \quad i=1,\dots,n,
\end{align}
where $q = 1.1$ and $b$ is such that $\sum_{i=1}^{n}\mu_i = 4$. Since $\lambda_i$ in (\ref{lambda_i}) decreases with $i$, people with lower indices get infected more quickly compared to people with higher indices. Since $\mu_i$ in (\ref{alpha_i}) increases with $i$, people with higher indices recover more quickly compared to people with lower indices. Thus, low index people get infected quickly and get well slowly.          

\begin{figure}[t]
	\begin{center}
	\subfigure[]{%
	\includegraphics[width=0.49\linewidth]{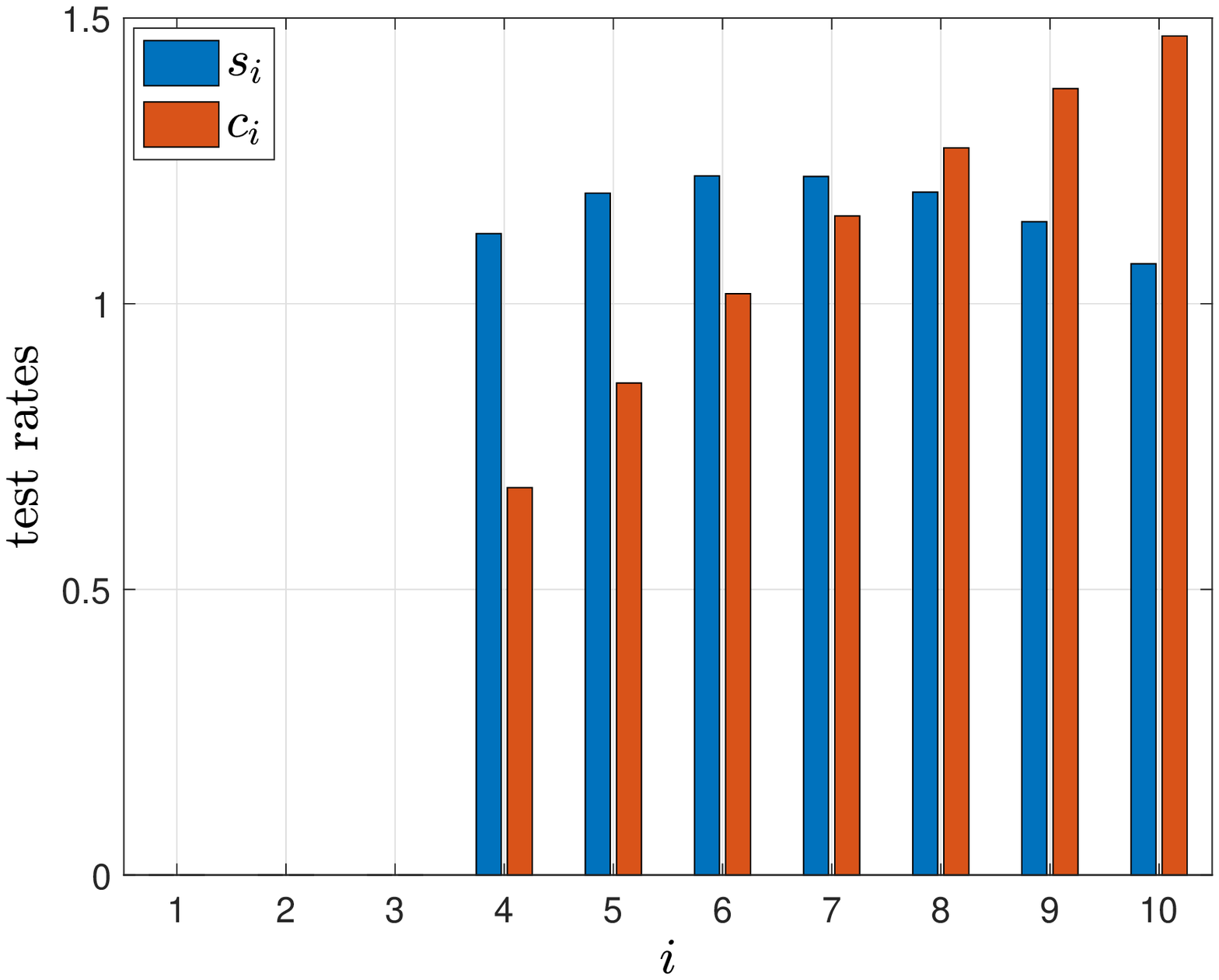}}
	\subfigure[]{%
	\includegraphics[width=0.49\linewidth]{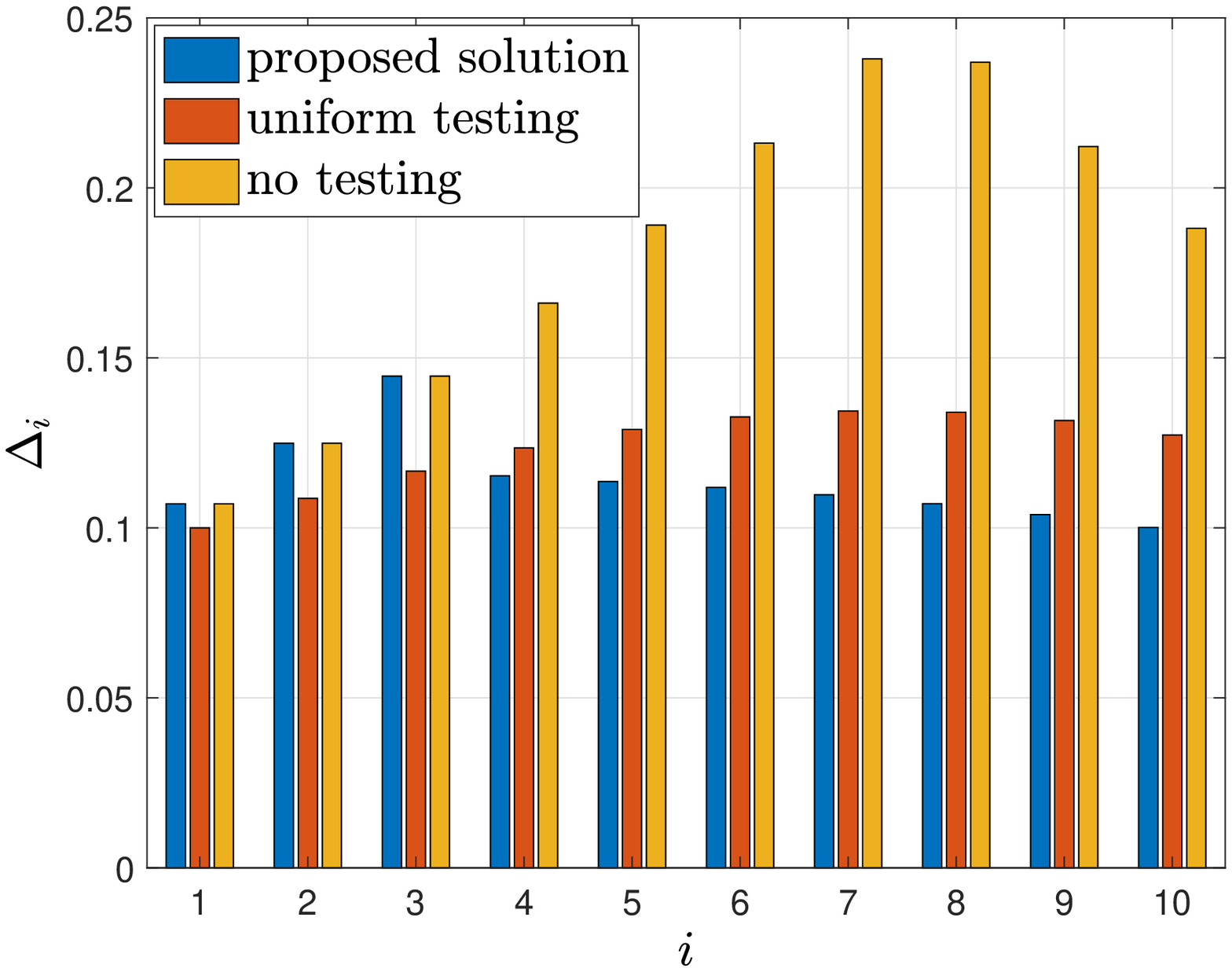}}
	\end{center}
	\caption{(a) Test rates $s_i$ and $c_i$, (b) corresponding average difference $\Delta_i$. }
	\label{Fig:sim1}
\end{figure}

In the first example, we take the total number of people as $n=10$, the total test rate as $C=16$, and $\theta = 0.5$. We start with randomly chosen $s_i$ and $c_i$ such that $\sum_{i=1}^{n}s_i+c_i =16$, and apply the alternating minimization based method proposed in Section~\ref{sect:opt_soln}. We repeat this process for 30 different initial $(s_i,c_i)$ pairs and choose the solution that gives the smallest $\Delta$. In Fig.~\ref{Fig:sim1}(a), we observe that the first three people are never tested by the health care provider. We note that $s_i$, which is the test rate when $\hat{x}_{i}(t)=0$,  initially increases with $i$ but then decreases with $i$. This means that people who get infected rarely are tested less frequently when they are marked as healthy. Similarly, we observe in Fig.~\ref{Fig:sim1}(a) that $c_i$, which is the test rate when $\hat{x}_{i}(t)=1$, monotonically increases with $i$. In other words, people who recover from the virus quickly are tested more frequently when they are marked infected.

In Fig.~\ref{Fig:sim1}(b), we plot $\Delta_i$ resulting from the solution found from the proposed algorithm, $\Delta_i$ when the health care provider applies tests to everyone in the population uniformly, i.e., $s_i = c_i=\frac{C}{2n}$  for all $i$, and $\Delta_i$ when the health care provider applies no tests, i.e., $s_i = c_i= 0$ for all $i$. In the case of no tests, we have $\Delta_i = \min\{ \frac{\theta \lambda_i}{\mu_i+\lambda_i}, \frac{(1-\theta) \mu_i}{\mu_i+\lambda_i}\}$. We observe in Fig.~\ref{Fig:sim1}(b) that the health care provider applies tests on people whose $\Delta_i$ can be reduced the most as opposed to uniform testing where everyone is tested equally. Thus, the first three people who have the smallest $\Delta_i$ are not tested by the health care provider. With the proposed solution, by not testing the first three people, $\Delta_i$ are further reduced for the remaining people compared to uniform testing. For the people who are not tested, the health care provider chooses $\hat{x}_i(t) = 1$ all the time, i.e., marks these people always sick as $\frac{\theta \lambda_i}{\mu_i+\lambda_i}> \frac{(1-\theta) \mu_i}{\mu_i+\lambda_i}$. This is expected as these people have high $\lambda_i$ and low $\mu_i$, i.e., they are infected easily and they stay sick for a long time. 

\begin{figure}[t]
	\centerline{\includegraphics[width=0.65\columnwidth]{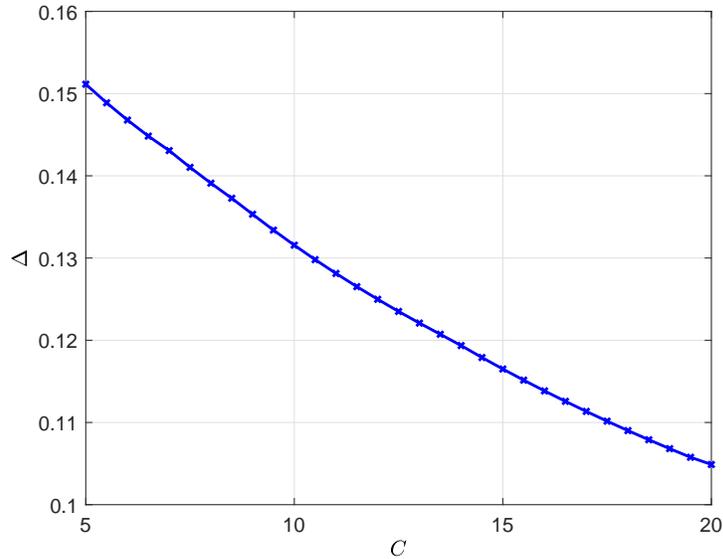}}
	\caption{The average difference $\Delta$ with respect to total test rate $C$.}
	\label{Fig:sim2}
\end{figure}

\begin{figure}[t]
	\centerline{\includegraphics[width=0.65\columnwidth]{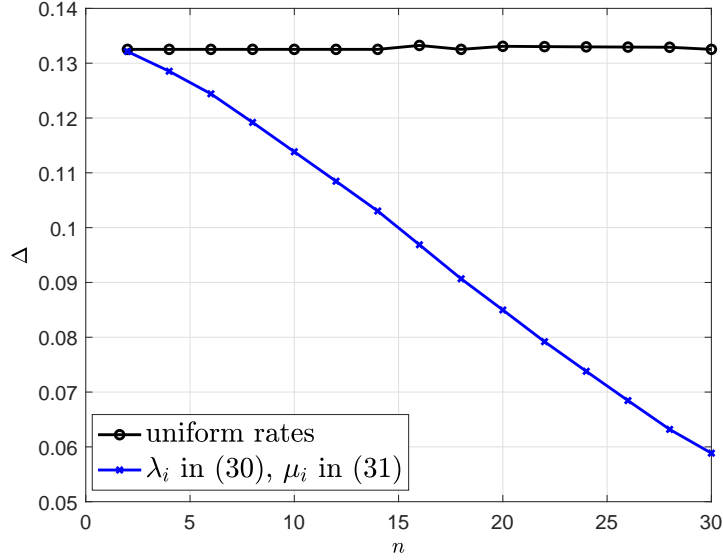}}
	\caption{The average difference $\Delta$ with respect to number of people $n$. We use uniform infection and healing rates, i.e., $\lambda_i = \frac{6}{n}$ and $\mu_i = \frac{4}{n}$ for all $i$, and also $\lambda_i$ in (\ref{lambda_i}) and $\mu_i$ in (\ref{alpha_i}) with $\sum_{i=1}^{n}\lambda_i = 6$ and $\sum_{i=1}^{n}\mu_i = 4$.}
	\label{Fig:sim4}
\end{figure}

In the second example, we use the same set of variables except for the total test rate $C$. We vary the total test rate $C$ in between $5$ and $20$. We plot $\Delta$ with respect to $C$ in Fig.~\ref{Fig:sim2}. We observe that $\Delta$ decreases with $C$. Thus, with higher total test rates, the health care provider can tract the infection status of the population better as expected.   

\begin{figure}[t]
	\begin{center}
	\subfigure[]{%
	\includegraphics[width=0.49\linewidth]{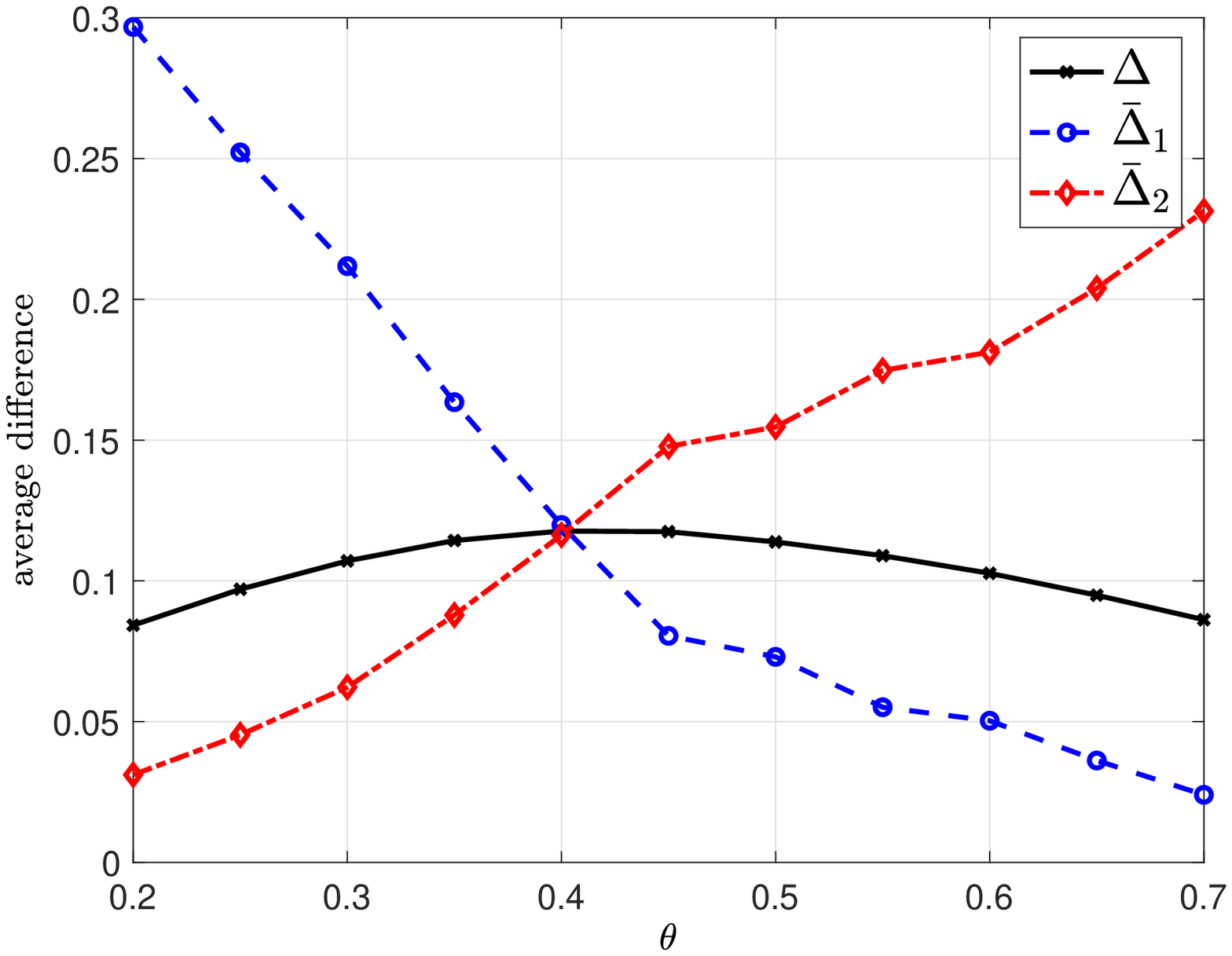}}
	\subfigure[]{%
	\includegraphics[width=0.49\linewidth]{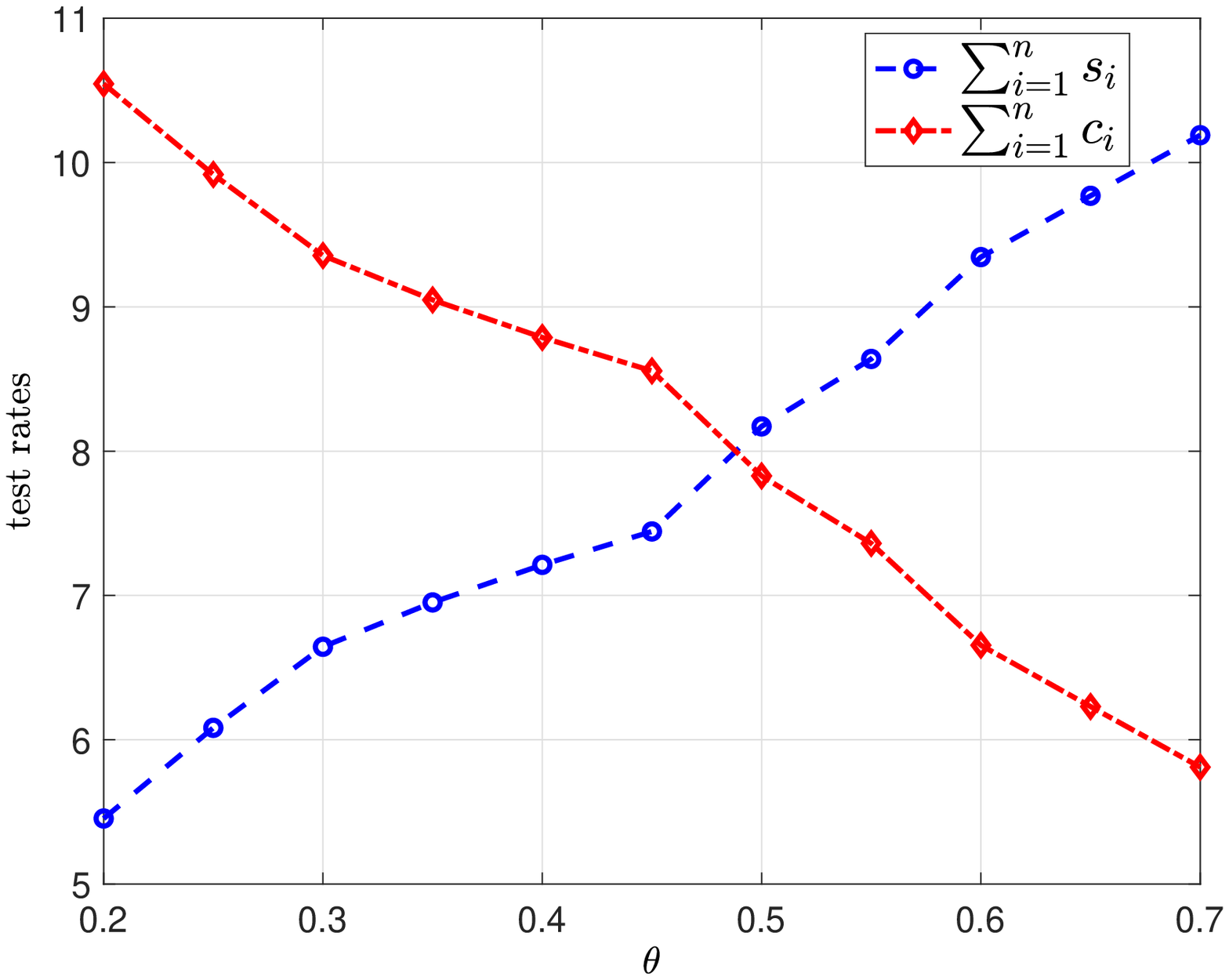}}
	\end{center}
	\caption{(a) $\Delta$ in (\ref{total_age}), $\bar{\Delta}_1$ which is  $\frac{1}{n}\sum_{i=1}^{n}\Delta_{i1}$, and $\bar{\Delta}_2$ which is  $\frac{1}{n}\sum_{i=1}^{n}\Delta_{i2}$, (b) corresponding total test rates $\sum_{i=1}^{n}s_{i}$ and $\sum_{i=1}^{n}c_{i}$. }
	\label{Fig:sim3}
\end{figure}

In the third example, we use the same set of variables except for the total number of people $n$. In addition, we also use uniform infection and healing rates, i.e., $\lambda_i = \frac{6}{n}$ and $\mu_i = \frac{4}{n}$ for all $i$, for comparison with $\lambda_i$ in (\ref{lambda_i}) and $\mu_i$ in (\ref{alpha_i}), while keeping the total infection and healing rates the same, i.e., $\sum_{i=1}^{n} \lambda_i= 6$ and $\sum_{i=1}^{n} \mu_i= 4$, for both cases. We vary the number of people $n$ from $2$ to $30$. We observe in Fig.~\ref{Fig:sim4} that when the infection and healing rates are uniform in the population, the health care provider can track the infection status with the same efficiency, even though the population size increases (while keeping the total infection and healing rates fixed). For the case of $\lambda_i$ in (\ref{lambda_i}) and $\mu_i$ in (\ref{alpha_i}), when we increase the population size, we increase the number of people who rarely get sick, i.e., people with high $i$ indices, and also people who rarely heal from the disease, i.e., people with small $i$ indices. Thus, it gets easier for the health care provider to track the infection status of the people. That is why when we use $\lambda_i$ in (\ref{lambda_i}) and $\mu_i$ in (\ref{alpha_i}), we observe in Fig.~\ref{Fig:sim4} that the health care provider can track the infection status of the people better, even though the population size increases.        

In the fourth example, we use the same set of variables as the first example except for the importance factor $\theta$. Here, we vary $\theta$ in between $0.2$ to $0.7$. We plot $\Delta$ in (\ref{total_age}), $\bar{\Delta}_{1}$ which is $\bar{\Delta}_1 = \frac{1}{n}\sum_{i=1}^{n}\Delta_{i1}$, and $\bar{\Delta}_{2}$ which is $\bar{\Delta}_2 = \frac{1}{n}\sum_{i=1}^{n}\Delta_{i2}$ in Fig.~\ref{Fig:sim3}(a). Note that $\bar{\Delta}_{1}$ represents the average difference when people are infected, but they have not been detected by the health care provider, and $\bar{\Delta}_2$ represents the average difference when people have recovered, but the health care provider still marks them as infected. Note that when $\theta$ is high, we give importance to minimization of $\bar{\Delta}_1$, i.e., the early detection of people with infection, and when $\theta$ is low, we give importance to minimization of $\bar{\Delta}_2$, i.e., the early detection of people who recovered from the disease. That is why we observe in Fig.~\ref{Fig:sim3}(a) that $\bar{\Delta}_1$ decreases with $\theta$ while $\bar{\Delta}_2$ increases with $\theta$. 

We plot the total test rates $\sum_{i=1}^{n}s_i$ and $\sum_{i=1}^{n}c_i$ in Fig.~\ref{Fig:sim3}(b). We observe in Fig.~\ref{Fig:sim3}(b) that if it is more important to detect the infected people, i.e., if $\theta$ is high, then the health care provider should apply higher test rates to people who are marked as healthy. In other words, $\sum_{i=1}^{n}s_i$ increases with $\theta$. Similarly, if it is more important to detect people who recovered from the disease, then the health care provider should apply high test rates to people who are marked as infected. That is, $\sum_{i=1}^{n}c_i$ is high when $\theta$ is low. Therefore, depending on the priorities of the health care provider, a suitable $\theta$ needs to be chosen.  

\begin{figure}[t]
	\begin{center}
	\subfigure[]{%
	\includegraphics[width=0.49\linewidth]{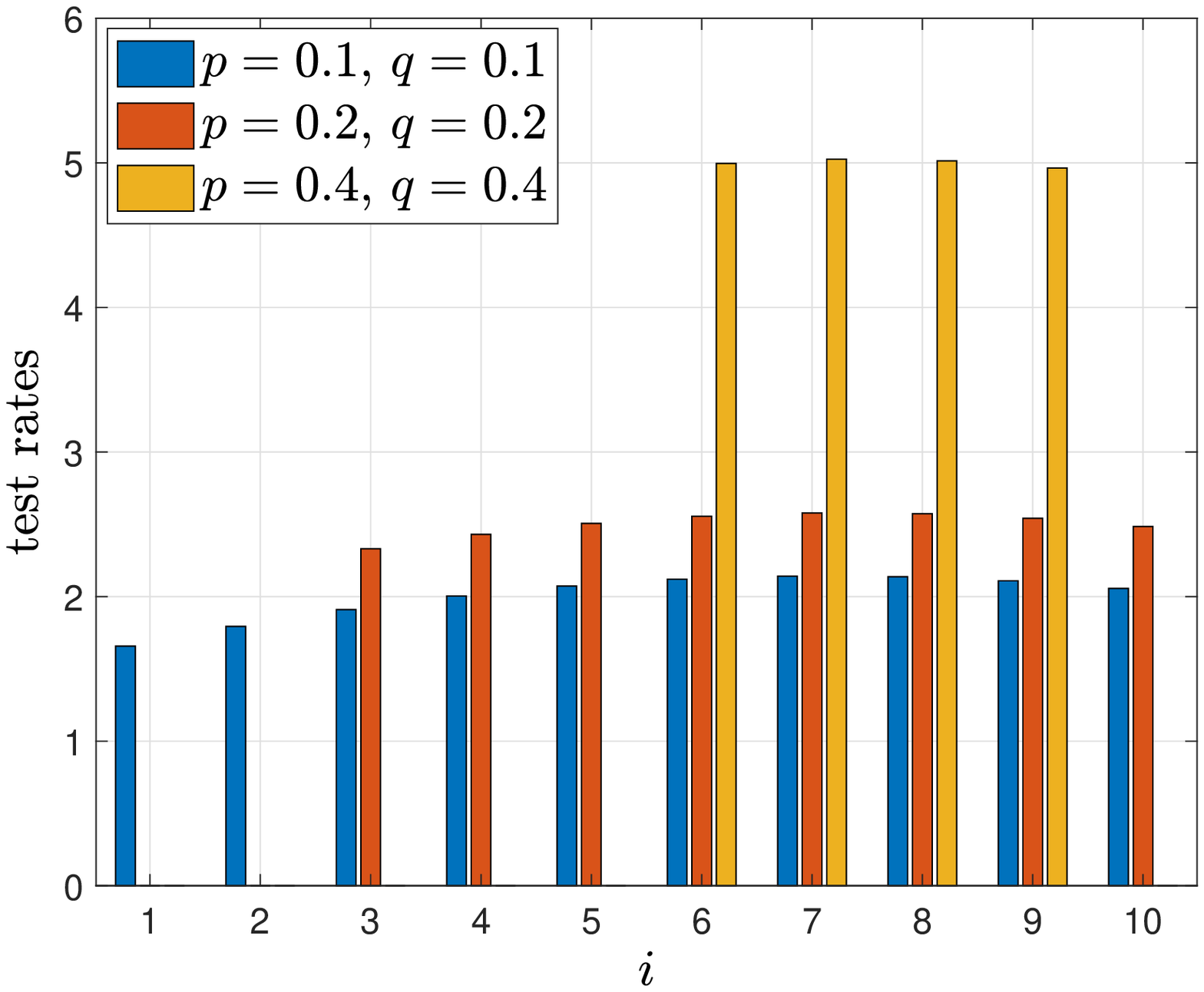}}
	\subfigure[]{%
	\includegraphics[width=0.49\linewidth]{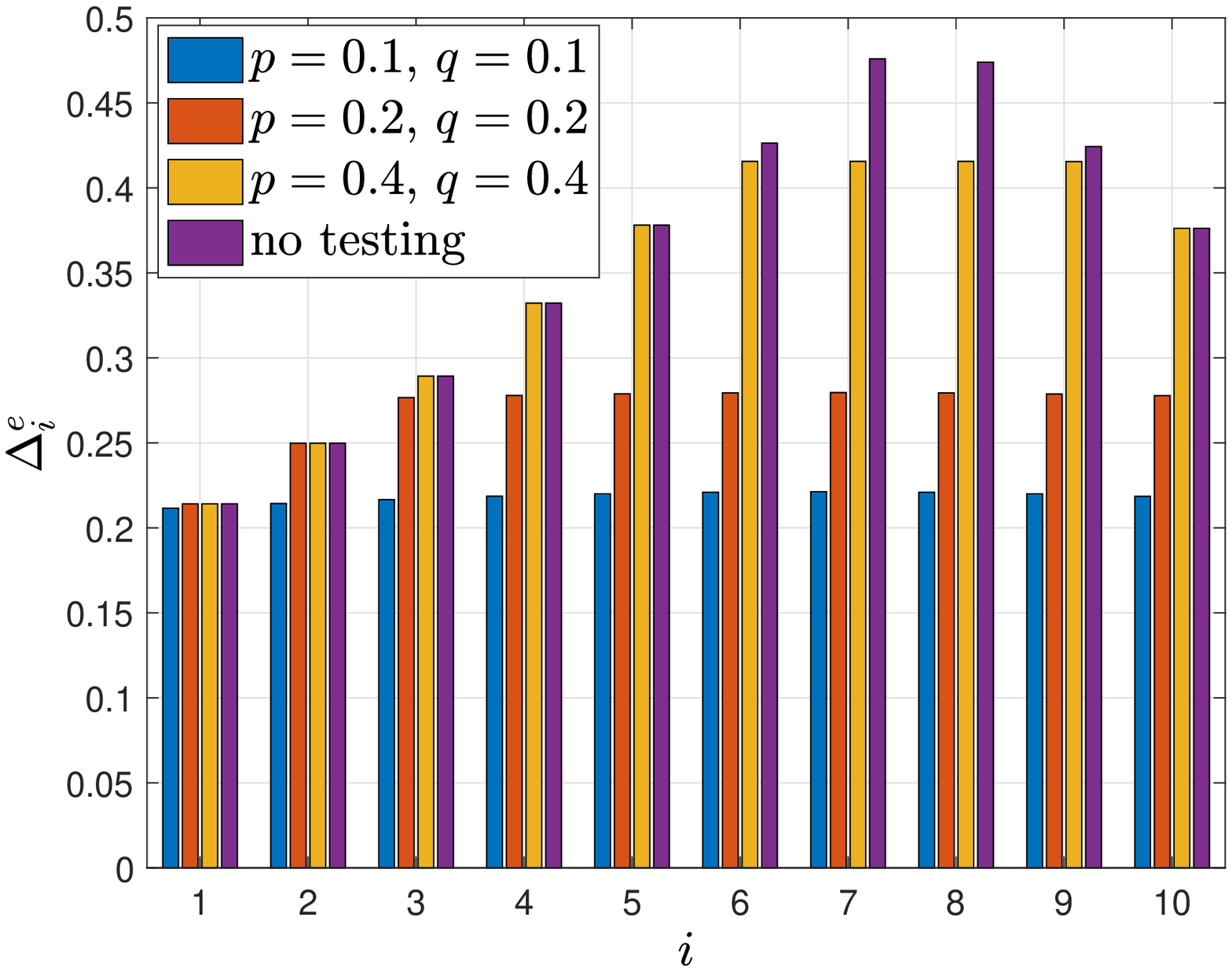}}
	\end{center}
	\caption{(a) Test rates $v_i$, (b) corresponding average difference $\Delta_{i}^e$ when there is error on the tests. }
	\label{Fig:sim5}
\end{figure}

In the fifth numerical result, we consider the case where there are errors in the test measurements, i.e., the model in Section~\ref{sect:error}. We take the total test rate as $C = 20$, and vary error rates in the test $p =q = \{0.1, 0.2, 0.4\}$. In Fig.~\ref{Fig:sim5}(a), we provide the test rates $v_i$ that we found as a result of our greedy policy in Section~\ref{sect:error}. When the error rates $p$ and $q$ are low, i.e., when $p =q = 0.1$, we see that the health care provider applies tests to everyone in the population and the corresponding $\Delta_{i}^e$ is lower than applying no test as shown in Fig.~\ref{Fig:sim5}(b). As we increase the error rates, we see that some people in the population start to be not tested by the health care provider, see Fig.~\ref{Fig:sim5}(a) when $p =q = \{0.2, 0.4\}$. In this case, the health care provider applies more tests to remaining people to combat with the test errors. However, even though it applies more tests to the remaining people, we observe in Fig.~\ref{Fig:sim5}(b) that the achieved average difference $\Delta_{i}^e$ gets higher as error rates increase.        

\begin{figure}[t]
	\begin{center}
	\subfigure[]{%
	\includegraphics[width=0.49\linewidth]{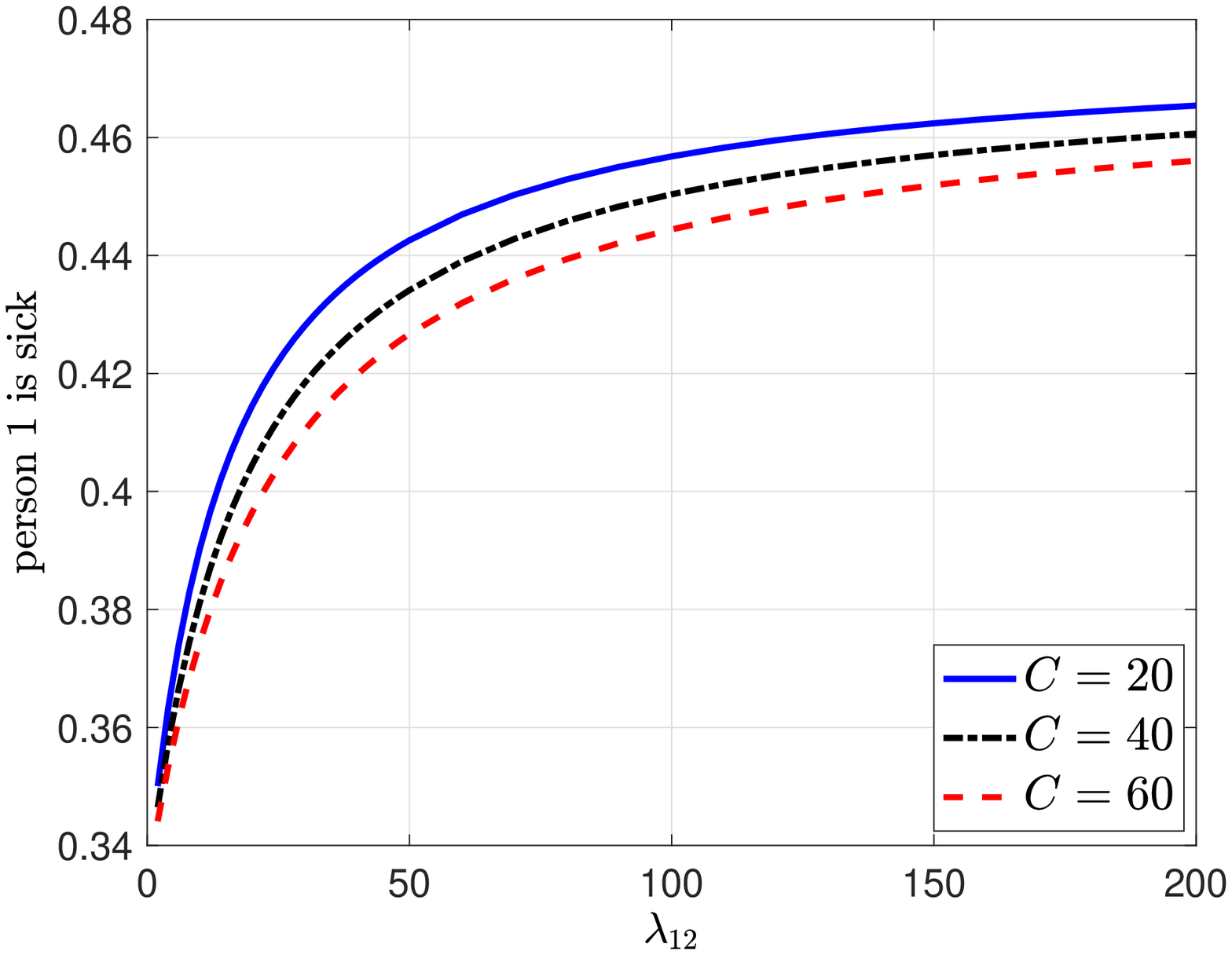}}
	\subfigure[]{%
	\includegraphics[width=0.49\linewidth]{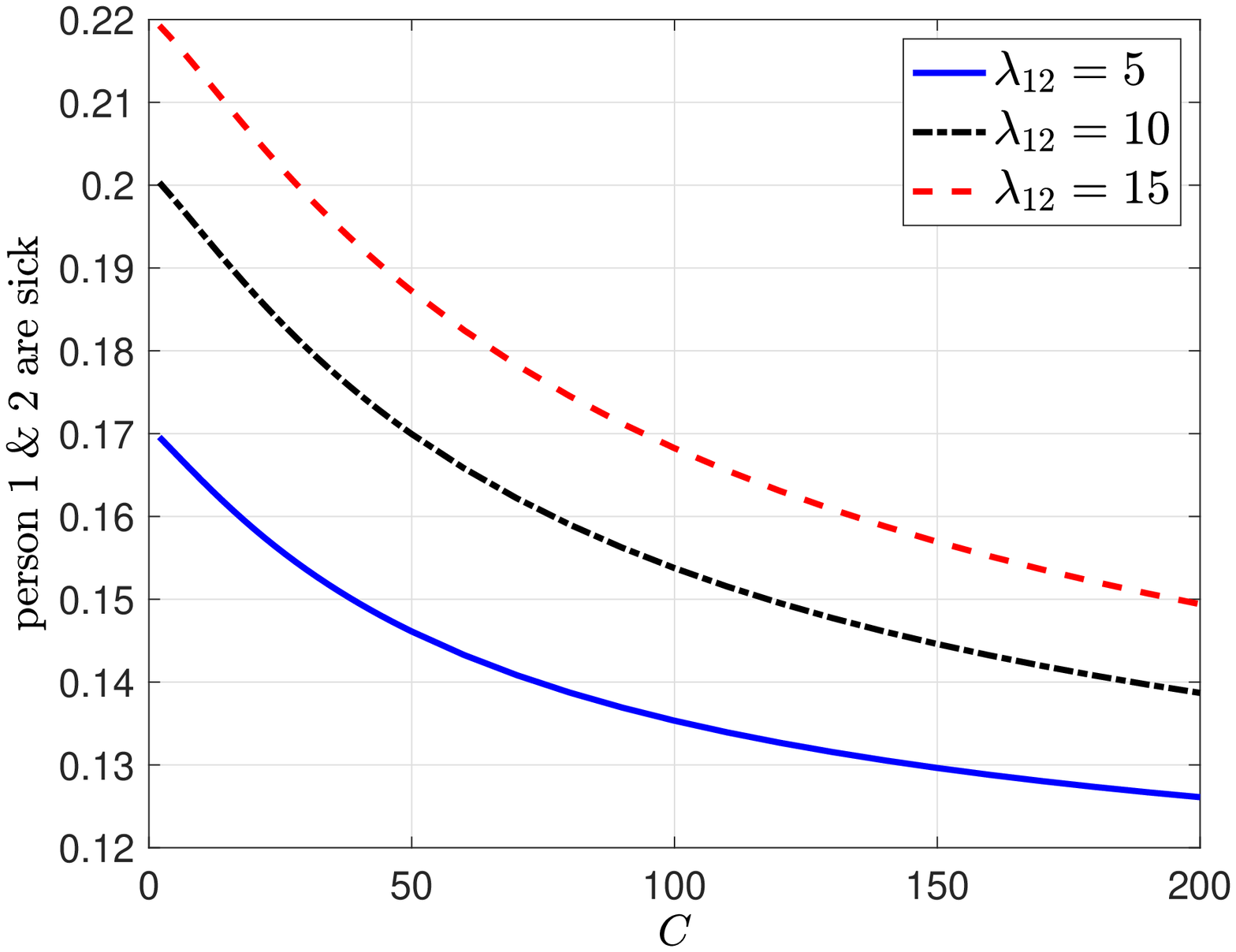}}
	\end{center}
	\caption{(a) The percentage of the time that person 1 stays as infected as we increase $\lambda_{12}$, (b) the percentage of the time that both person 1 and 2 stay as infected as we increase the total test rate $C$. }
	\label{Fig:sim7}
\end{figure}

In the sixth numerical result, we consider the case where the infection status of the people depend on each other. In other words, when one person is infected, it can infect the other person with rate $\lambda_{12}$ when it is not detected by the health care provider, i.e., the infection model in Section~\ref{sect:Dependent}. For this example, first, we take $\mu =5$, $\lambda= 2.5$, $s=c = \frac{C}{4}$ and vary $\lambda  = \{2, \dots, 200\}$ and $C = \{20, 40, 60\}$. If $\lambda_{12} =0$, i.e., if the infection status of people are independent from each other, then the average time that person 1 or 2 is sick is equal to $\frac{\lambda}{\lambda+\mu} = \frac{1}{3}$. As we increase infection rate $\lambda_{12}$ among the person 1 and 2, we see in Fig.~\ref{Fig:sim7}(a) that the average time that person 1 is sick increases. However, we note that as we increase the total test rate, the health care provider can detect a sick person more frequently, and that is why the average infected time is low in Fig.~\ref{Fig:sim7}(a) when the test rate is high. Then, we consider $\lambda_{12} = \{5,10,15\}$ and vary the total test rates $\lambda  = \{2, \dots, 200\}$. We plot the average time that both person 1 and 2 stay as sick in Fig.~\ref{Fig:sim7}(b). As we increase the total test rate, the health care provider detects the infected person more quickly, and thus, prohibits the infection from spreading. As a result, we observe that the average time that both people are infected decreases in $C$ in Fig.~\ref{Fig:sim7}(b). Since both people can be infected with the virus independent from each other with rate $\lambda$, the plots in Fig.~\ref{Fig:sim7}(b) do not go down to 0.    

\begin{figure}[t]
	\begin{center}
	\subfigure[]{%
	\includegraphics[width=0.49\linewidth]{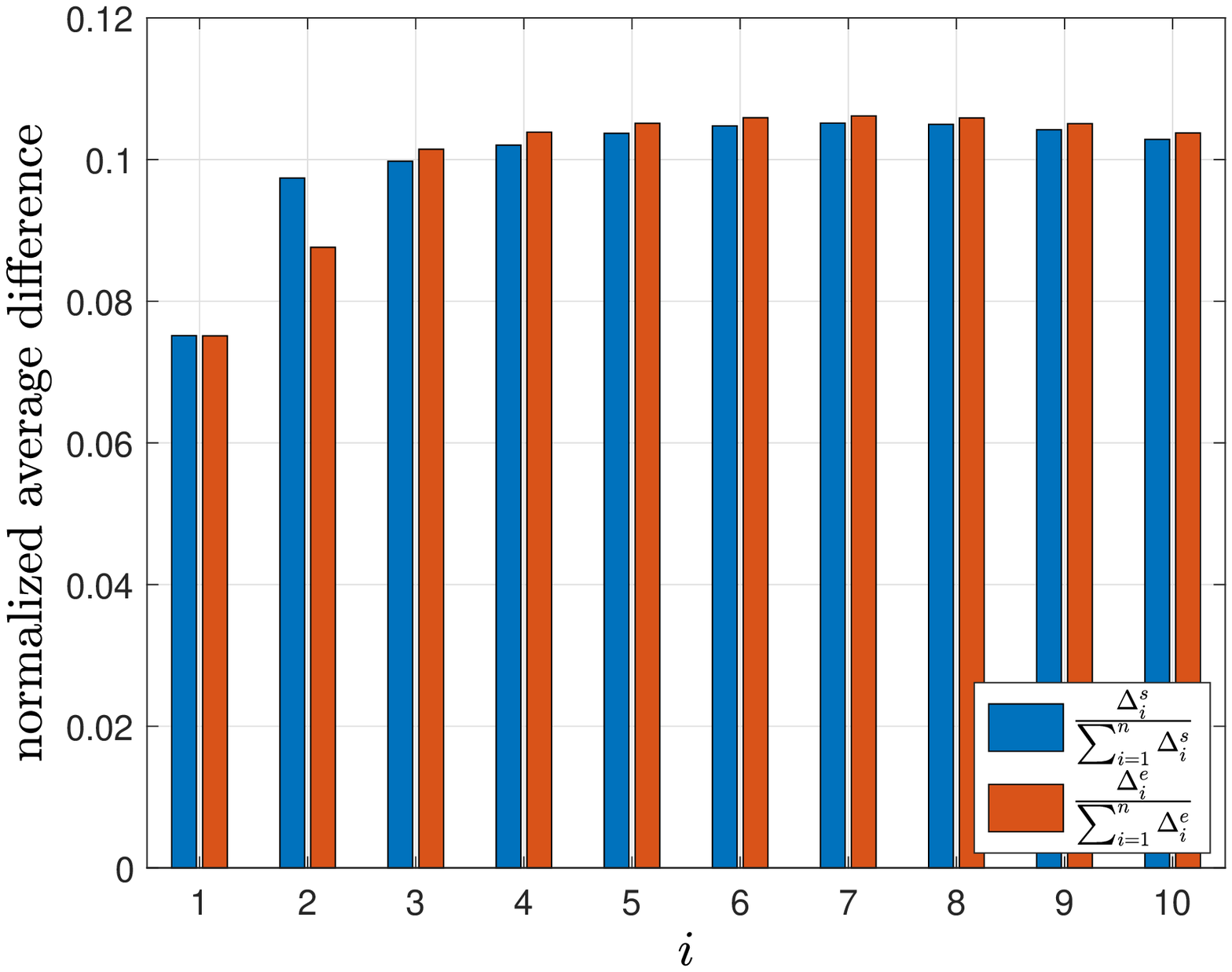}}
	\subfigure[]{%
	\includegraphics[width=0.49\linewidth]{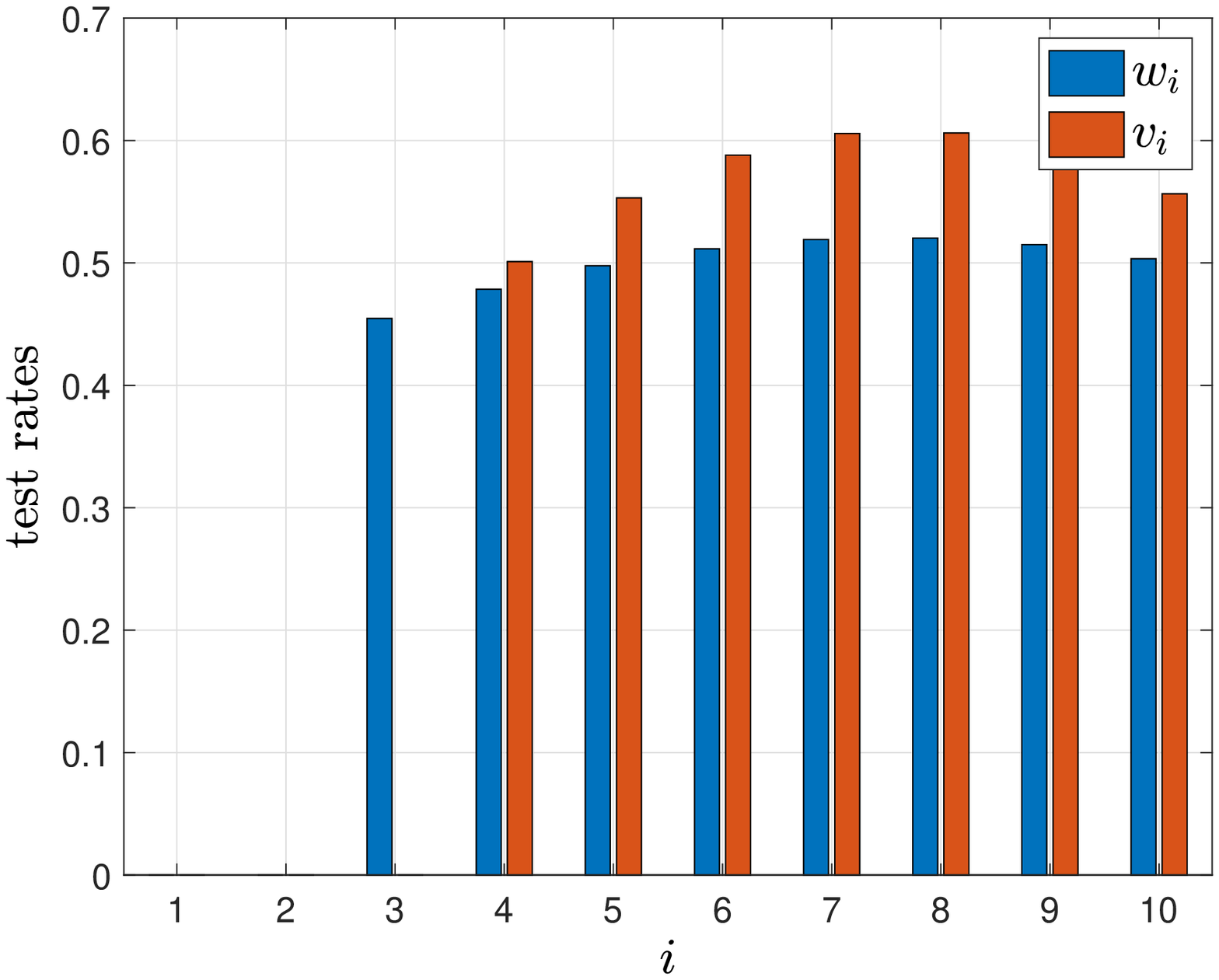}}
	\end{center}
	\caption{(a) The normalized average differences $\frac{\Delta_{i}^s}{\sum_{i=1}^{n}\Delta_{i}^s}$, and $\frac{\Delta_{i}^e}{\sum_{i=1}^{n}\Delta_{i}^e} $, and (b) the corresponding test rates $w_i$ and $v_i$. }
	\label{Fig:sim8}
\end{figure}

In the last numerical result, we consider the age of incorrect information based error metric in Section~\ref{sect:aoii_model}. Here, the estimation error increases with the time that the health care provider does not detect the changes in the infection status of the people. As a result, the average difference expression given by $\Delta_{i}^s$ in (\ref{eqn_aoii_final}) is different than $\Delta_{i}^e$ in (\ref{Delta_ei_val}) when $p=q=0$. For this example, we consider the total test rate $C=4$ and compare the normalized average differences given by $\frac{\Delta_{i}^s}{\sum_{i=1}^{n}\Delta_{i}^s}$, and $\frac{\Delta_{i}^e}{\sum_{i=1}^{n}\Delta_{i}^e} $ and the corresponding test rates $w_i$ and $v_i$. In Fig.~\ref{Fig:sim8}(b), depending on the error metric model, people who are tested by the health care provider and their test rates vary considerably. For example, with the error metric $\Delta_{i}^s$ in (\ref{eqn_aoii_final}), we apply tests to the 3rd person while the same person is not tested with the error metric $\Delta_{i}^e$ in (\ref{Delta_ei_val}). In Fig.~\ref{Fig:sim8}(a), we provide the normalized average difference values. Here, the average normalized error for the tested people has similar values whereas the normalized difference may vary for the untested people. Thus, we should choose a suitable error metric that satisfies the priorities of the health care provider as it greatly affects who is tested and with which test rates.          

\section{Conclusion and Discussion}
We considered timely tracking of infection status of individuals in a population. For exponential infection and healing processes with given rates, we determined the rates of exponential testing processes. We considered errors on the test measurements and observed that in order to combat the test errors, a limited portion of the population may be tested with higher test rates. Then, we studied a dependent infection spread model for two people where an infected person can spread the virus to the other one if it has not been detected by the health care provider yet. Finally, we studied an AoII based error metric where the error function linearly increases over time as the changes on the infection status has not been detected by the health care provider. We observed in numerical results that the test rates depend on the individuals' infection and healing rates, the individuals' last known state of healthy or infected, as well as the health care provider's priorities of detecting infected people versus detecting recovered people more quickly.

In the literature, in order to model epidemics, population is partitioned into groups called \textit{compartments}. One such example is the SIR model used in \cite{SIR_model} with the compartments susceptible (S), infected (I), and recovered (R) which has been further developed by adding states hospitalized (H), and death (D) in \cite{yagiz2020}. In these epidemic models, the transitions between the compartments are assumed to be Markovian. In \cite{yagiz2020}, with the epidemiological data, the delay distributions for the infected (I) to hospitalized (H), and infected (I) to death (D) are well approximated by exponential and gamma distributions, respectively. However, due to the lack of data availability the delay distribution for infected (I) to recovered (R) is modeled with gamma distribution with higher tolerance. In our work, we modeled infection and recovery times, i.e., the delays between recovered (R) to infected (I) and infected (I) to recovered (R) with exponential distributions. Therefore, more realistic infection tracking models can be developed by considering gamma distributions as observed in \cite{yagiz2020}. This more realistic model corresponds to the problem of real-time timely tracking of a binary Markov source in a serially connected network. The serially connected network model has been studied in \cite{Yates20} with the traditional age of information metric. We note that considering the same networking model with the AoII based error metric to track information dissemination of a binary Markov source is a promising research direction and has direct applications to the real-time tracking of epidemic spread models. One can also study the extension of dependent infection spread model in Section~\ref{sect:Dependent} to $n>2$ people as a future research direction.

\bibliographystyle{ieeetr}
\bibliography{IEEEabrv,myLibrary_bastopcu}
\end{document}